\documentclass[10pt,letterpaper]{article}
\usepackage[top=0.85in,left=1.5in,footskip=0.75in,marginparwidth=2in]{geometry}

\usepackage[utf8]{inputenc}

\usepackage{cite}

\usepackage{nameref,hyperref}

\usepackage{microtype}
\DisableLigatures[f]{encoding = *, family = * }

\raggedright
\usepackage{changepage}

\usepackage[aboveskip=1pt,labelfont=bf,labelsep=period,singlelinecheck=off]{caption}

\makeatletter
\renewcommand{\@biblabel}[1]{\quad#1.}
\makeatother

\usepackage{lastpage,fancyhdr,graphicx}
\usepackage{epstopdf}
\fancyheadoffset[L]{2.25in}
\fancyfootoffset[L]{2.25in}

\usepackage{color}

\definecolor{Gray}{gray}{.25}

\usepackage{graphicx}

\usepackage{sidecap}
\usepackage{algorithmic}
\usepackage{algorithm}
\usepackage{amsmath,amssymb,amsfonts}
\usepackage{subfigure}
\usepackage{wrapfig}
\usepackage[pscoord]{eso-pic}
\usepackage[fulladjust]{marginnote}
\reversemarginpar

\begin{document}
\vspace*{0.35in}

\begin{flushleft}
{\Large
\textbf\newline{Africa 3: A Continental Network Model to Enable the African Fourth Industrial Revolution.}
}
\newline
\\
 
Olasupo O. Ajayi\textsuperscript{1},
Antoine B. Bagula\textsuperscript{1},
Hloniphani M. Maluleke \textsuperscript{1}
\\
\bigskip
\bf{1} Department of Computer Science,University of the Western Cape, Bellville, South Africa
\\

\bigskip
* olasupoajayi@gmail.com

\end{flushleft}

\begin{abstract} It is widely recognised that collaboration can help fast-track the development of countries in Africa. Leveraging on the fourth industrial revolution, Africa can achieve accelerated development in health care services, educational systems and socio-economic infrastructures. While a number of conceptual frameworks have been proposed for the African continent, many have discounted the Cloud infrastructure used for data storage and processing as well as the underlying network infrastructure upon which such frameworks would be built. This work therefore presents a continental network model for interconnecting nations in Africa through its data centres. The proposed model is based on a multilayer network engineering approach, which first groups African countries into clusters of data centers using a hybrid combination of clustering techniques; then utilizes Ant Colony Optimisation with Stench Pheromone, that is modified to support variable evaporation rates, to find ideal network path(s) across the clusters and the continent as a whole. 
The proposed model takes into consideration the geo-spatial location, population sizes, data centre counts and intercontinental submarine cable landings of each African country, when clustering and routing. For bench-marking purposes, the path selection algorithm was tested on both the obtained 
clusters and African Union’s regional clusters.

\end{abstract}

Africa, Ant Colony Optimization, Clustering Algorithms, Communication Networks, Computer Networks 


\section{INTRODUCTION}

Despite the widespread adoption and progress made toward the Sustainable Development Goals (SDGs), Africa still lags behind most of the world when it comes to socio-economic and industrial development.  According to the
United Nations' report on global human development index, as of today, only seven (7) of Africa’s 55 countries have a Human Development Index (HDI) score greater than 0.75 \cite{uno}. The HDI is a metric used to measure the 
state of development of countries around the world. It takes into consideration factors such as level of healthcare, education, average life expectancy, among other things. It can clearly be inferred from this report that 75\% of African countries are still developing 
countries, with poor infrastructure for health, education, transportation, electricity etc. Collaboration has been identified as a potential medium for accelerating development in many of these countries. A number of such collaborative efforts for development of 
various aspects of Africa include: regional economic communities such as the Economic Community of West African States (ECOWAS) and Southern African Development Community (SADC), for economic and political developments \cite{AU}; research and education networks such as the West and Central African Research and Educational Network (WACREN) and UbuntuNet, for education \cite{AfREN} and communication networks such as the Eastern Africa Submarine Cable System (ESSAy) and Melting Pot 
Indian-oceanic Submarine System (METISS) \cite{sub}. Besides these efforts, researchers have also proposed numerous frameworks for collaboration and resource sharing in Africa, such as those in the works of \cite{FedCloud, Mandava, Adams12, chavula}. 
Despite these proposals, only a few works have focused on the underlying network architecture required to support such inter-continental collaborations. In \cite{africa1}, the authors considered an optical network to link Africa, while in \cite{africa2}, a dual ring 
network topology was proposed for interconnecting African countries. In \cite{djamen}, the role of the Internet on the evolution of communication infrastructure in 8 west African countries was considered. In \cite{is}, the author presented a holistic view of community networks in African countries, including the current state and various implementation barriers. 

With respect to Clustering Algorithms (CAs), the authors in \cite{c1} surveyed the state of the art, discussed common distance, similarity and evaluation functions of a number of CAs. They also presented different classes of traditional CAs, including: partition, hierarchical, fuzzy, distribution, density and graph; as well as modern CAs, some of which include: kernel based, swarm intelligence, affinity and hybrids. In \cite{c2}, Cheng, \textit{et al.} proposed a CA that is robust to noise. The algorithm was built upon a hierarchical model, with neighour density used to filter out the noise, while modified graph algorithms were used to form the required clusters. In \cite{c3}, an improvement on classic density-based CA was proposed to tackle the well-known issue of determining ideal scanning radius. The proposed model called DHeat incorporated an adaptive scanning radius mechanism. A number of other contemporary CAs exists but most are built upon the classic CAs, such as those in \cite{c4}, \cite{c5}. In this work, we would stick with the classic CAs and apply them to clustering the African continent.

It is widely recognised that collaboration and digitisation can help achieve the SDGs by building upon the technologies of the fourth industrial revolution (4IR) to get a connected continent with potential for collaboration in health care services, industry and education. It can also deliver an Africa where infrastructure are shares in a federated way to bridge the gap between the poorest and the richest countries on the continent. A Federated Cloud model for health care in Africa was proposed in~\cite{FedCloud} as an attempt to 
achieve such goal. The work in~\cite{FedCloud} investigated both cooperative and competitive Cloud models and tested five different workload allocation schemes. Results from that work, showed that allocation delays were lower and resources were better utilised in the cooperative model, while completion times and quality of service were better provided using the competitive model. Despite proposing a potential model, no discussion was made on the implementation of such a model, especially from a network 
architecture perspective. Prior to the work done in~\cite{FedCloud},  Huang, {\it et al.} had investigated the viability of a network, referred to as  ``Africa 2'', that covers the entire African continent \cite{africa2}. They proposed a dual-ring network 
architecture to service the 55 countries (54 plus Western Sahara*) of Africa \cite{AU}. In their work, they selected some countries as “bridge-nodes” with links between these nodes serving as connection points between the two rings. The work also compared a greedy algorithm and simulated annealing algorithm for selecting the optimal placement of the bridge node. Furthermore, they considered various models for achieving inter-connectivity among African countries, while factoring in growth potentials. 

This paper revisits the issue of African connectivity and proposes a continental network model called ``{\it Africa 3}'', that can enable the African fourth industrial revolution through digitisation. In this work, we seek to address similar routing objectives as 
``{\it Africa 2}''; however, by applying a completely different hierarchical approach. In our model, neighbouring countries are grouped into clusters forming digital islands. These clusters are then interconnected by a network engineered model based on a modified ant-colony optimisation algorithm. Africa 3, also borrows the federated approach from~\cite{FedCloud}, which enables efficient connection of Cloud infrastructures within the continent.

The specific contributions of this work are:
\begin{enumerate}
    \item Development of a continental network model for intercommunication across Africa. This model takes into consideration the technological capabilities (in terms of Cloud data centres), the population and geo-political importance of each African country.
    
    \item Division of African countries into functional clusters using a number of clustering algorithms. This enables easy “visualisation” of the influence each feature considered could have on the continent.
    
    \item The proposal of a modified Ant Colony Routing algorithm for the identification of the best network routes within and between clusters of countries across Africa. The algorithm is based a combination of dynamic stench pheromones and weighted cost function. 
 \end{enumerate}

With respect to the organization of this paper, following this section is section \ref{s2}, with a review of related literature and the introduction of our proposed continental network model for Africa. In section \ref{s3}, we consider various clustering algorithms for the lower layer of Africa 3, while network routing at the upper layer is presented in section \ref{s4}. Simulations and discussions of experimental results are done in section \ref{s5}, while open implementation challenges are highlighted in section \ref{s6}. We then conclude the paper and give insights into future research directions in section \ref{s7}.

\section{The African Network Model} \label{s2}
As suggested earlier, ``{\it Africa 3}''  is an continental network that combines the routing objectives of ``{\it Africa 2}''~\cite{africa2} with the Cloud federation approach proposed in~\cite{FedCloud}. Africa 3 consists of a hierarchical network, wherein 
at the lower level, neighbouring countries are grouped into clusters and the cluster heads or regional hubs are selected as the “bridge-node”. These bridge-nodes are then linked together, forming a continent-wide network at the higher level. The choice of which country is selected as a cluster head is based on population size, number of data centres, geographical location and regional importance. Table \ref{tab1} puts these in perspective and shows the number of data centres in African 
countries~\cite{dcmap}, as well as the approximate population size and centroids of each country \cite{afPop}. On Table~\ref{tab1}, data centres values of zero implies that the country either does not have any data centre or the ones present therein are below the standard stipulated in \cite{diminico}.

\begin{table*}[]
\centering
\caption{Geo-demography of African Countries}
\label{tab1}
\begin{scriptsize}
\begin{tabular}{l l c l l c}
\hline
\textbf{SN} & \textbf{Country} & \textbf{Sub-Region} & \textbf{Centroid Coordinates} & \textbf{Pop. Est.} & \textbf{DC Count}
\\ \hline
1 &	Burundi &	Central Africa & 	-3.35939665747, \par 29.8751215645 &	9,823,828 &	0 \\ \hline

2 &	Cameroon &	Central Africa &	5.69109848986, \par 12.7396415575 &	21,917,602 &	0 \\ \hline
3 &	Central African Republic &	Central Africa &	6.56823297048, \par 20.4682683098 &	3,859,139	 & 0 \\ \hline
4 &	Chad &	Central Africa &	15.333337585, \par 18.6449251291	& 11,039,873	& 0 \\ \hline
5	& Democratic Republic of Congo &	Central Africa &	-2.8774628897, \par 23.643961066 &	86,026,000 &	1 \\ \hline
6	& Equatorial Guinea	& Central Africa &	1.70555135464, \par 10.3413792389 &	1,222,442 &	0 \\ \hline
7	& Gabonese Republic &	Central Africa &	-0.586600249551, \par 11.7886286968 &	1,802,278	& 0 \\ \hline
8 &	Republic of the Congo &	Central Africa &	-0.837874631184, \par 15.2196576195 &	3,697,490 &	0 \\ \hline
9 &	Sao Tome and Principe &	Central Africa &	0.443914449308, \par 6.72429657927 &	179,200	 & 0 \\ \hline

10	& Comoros &	Eastern Africa &	-11.8778344355, \par 43.6825396848 &	806,200	 & 0 \\ \hline
11 &	Djibouti &	Eastern Africa &	11.7487180559, \par 42.5606754026 &	864,618	& 0 \\ \hline
12	& Eritrea &	Eastern Africa &	15.361866184, \par 38.8461701125 &	6,536,000 &	0 \\ \hline
13 &	Ethiopia &	Eastern Africa &	8.6227867931, \par 39.6008009763 & 112,078,730	& 0 \\ \hline
14	& Kenya &	Eastern Africa &	0.599880215445, \par 37.7959397293 & 45,533,000	& 6 \\ \hline
15 & Madagascar	& Eastern Africa &	-19.3718958677, \par 46.7047367428 & 22,434,363	& 1 \\ \hline
16 &	Mauritius &	Eastern Africa &	-20.2776870433, \par 57.5712055061 &	1,261,208 &	9 \\ \hline
17	& Rwanda	& Eastern Africa &	-1.99033831693, \par 29.9198851524	& 10,515,973	& 0 \\ \hline
18 &	Seychelles &	Eastern Africa &	-4.66099093522, \par 55.4760327912	& 90,945 &	0 \\ \hline
19 &	Somalia	& Eastern Africa &	4.75062876, \par 45.70714487	& 12,316,895 & 0 \\ \hline
20 &	South Sudan &	Eastern Africa &	7.30877944922, \par 30.2479000186	& 8,260,490	& 0 \\ \hline
21 & Tanzania	& Eastern Africa &	-6.27565408332, \par 34.8130998093	& 51,046,000	& 1 \\ \hline
22	& Uganda &	Eastern Africa &	1.27469298731, \par 32.3690797137	& 34,856,813	& 0 \\ \hline

23 &	Algeria	& Northern Africa &	28.1589384945, 2.6173230092	& 40,100,000	& 1 \\ \hline 

24 &	Egypt &	Northern Africa &	26.4959331064, \par 29.8619009908	& 98,002,045	& 11 \\ \hline 
25 &	Libya	& Northern Africa &	27.0309449491, \par 18.0086616872	& 5,298,152	& 0 \\ \hline 
26 &	Morocco	& Northern Africa &	29.8376295469, \par -8.45615794798 &	33,337,529 &	5 \\ \hline 
27 &	Sudan &	Northern Africa &	15.9903566856, \par 29.9404681199	& 40,235,000 & 0 \\ \hline 
28 &	Tunisia &	Northern Africa &	34.1195624649, \par 9.55288358933 &	10,982,754	& 2 \\ \hline 
29 &	Western Sahara*	& Northern Africa &	24.2295673865, \par -12.2198275485	& 510,713 &	0 \\ \hline 
30	& Mauritania &	Northern Africa &	20.257367056, \par -10.3477981455 &	3,718,678 &	0 \\ \hline 

31 &	Malawi &	Southern Africa &	-13.2180808806, \par 34.2893559871 &	16,832,900	& 0 \\ \hline 
32	& Mozambique &	Southern Africa &	-17.2738164259, \par 35.5336754259 &	28,013,000	& 0 \\ \hline 
33 &	Zambia &	Southern Africa &	-13.4582415186, \par 27.7747594637 &	15,473,905	& 0 \\ \hline 
34 &	Zimbabwe &	Southern Africa &	-19.0042041882, \par 29.8514412019	& 13,061,239 &	1 \\ \hline 
35 &	Angola &	Southern Africa &	-12.2933605438, \par 17.5373676815 &	24,383,301	& 3 \\ \hline 
36 &	Botswana &	Southern Africa &	-22.1840321328, \par 23.7985336773	& 2,024,904	& 0 \\ \hline 
37	& Eswatini & 	Southern Africa &	-26.5584304502, \par 31.4819369011	& 1,119,375	& 0 \\ \hline 
38	& Lesotho &	Southern Africa &	-29.5800318833,\par 28.2272313098	& 1,894,194	& 0 \\ \hline 
39	& Namibia &	Southern Africa &	-22.1303256842,\par 17.209635667	& 2,280,700	& 0 \\ \hline 
40	& South Africa &	Southern Africa &	-29.0003409534,\par 25.0839009251	& 58,558,270 &	21
\\ \hline

41	& Benin &	Western Africa &	9.64175970225, \par 2.32785254449	& 10,008,749 &	0 \\ \hline 
42	& Burkina Faso &	Western Africa &	12.2695384574, \par -1.75456600914 &	18,450,494	& 0 \\ \hline 
43 &	Cabo Verde &	Western Africa &	15.9552332421, \par -23.9598882026	& 491,875	& 0 \\ \hline 
44	& Gambia &	Western Africa &	13.4496524352, \par -15.3960129463	& 1,882,450	& 0 \\ \hline 
45 &	Ghana &	Western Africa &	7.95345643541, \par -1.21676565807 &	27,043,093	& 2\\ \hline 
46	& Guinea &	Western Africa &	10.4362159331, \par  -10.9406661161 &	10,628,972	& 0 \\ \hline 
47 &	Guinea Bissau &	Western Africa &	12.0474494815, \par -14.9497244459 & 1,530,673 & 0 \\ \hline 
48 &	Cote d'Ivoire &	Western Africa &	7.62842620235,\par -5.5692156998 &	22,671,331	& 0 \\ \hline 
49	& Liberia &	Western Africa &	6.45278491657,\par -9.3220757269 &	3,476,608	& 0 \\ \hline 
50 &	Mali &	Western Africa &	17.3458158135, \par -3.54269064851 &	14,528,662	& 0 \\ \hline 
51 &	Niger &	Western Africa &	17.4191249296, \par 9.38545881539 & 17,138,707	& 0 \\ \hline 
52 &	Nigeria &	Western Africa &	9.59411452233, \par 8.08943894771 &	200,963,599	& 10 \\ \hline 
53 &	Senegal	& Western Africa &	14.3662417313, \par -14.4734923973	& 14,354,690	& 0 \\ \hline 
54 &	Sierra Leone &	Western Africa &	8.56329593038, \par -11.7927124668 &	6,348,350	& 0 \\ \hline 
55	& Togo &	Western Africa &	8.52531356141, \par 0.962328449769 &	6,191,155 &	0 \\ \hline 

\end{tabular}
\end{scriptsize}
\end{table*}

Finding  the optimal or “best” path in a network environment has been found to be  an NP-hard problem in many cases and one which cannot be solved in intrinsic computing time. A number of meta-heuristic approaches have been developed to solve the 
optimal path finding problem, prominent among which is the Ant Colony Optimization (ACO). Ant Colony Optimization (ACO) is a meta-heuristic algorithm that draws inspiration from the self-organizing behaviour of ants \cite{aco}. Like in real ants which 
randomly search for food and lay pheromones to guide other ants to food sources, ‘ants’ in ACO also update pheromone levels along traversed pathways. These artificial ants have memories, where they keep information about previously visited paths and local solution. Upon returning to the ‘nest’ from food foraging, each ant updates the pheromone level on the trail it traversed. These pheromone levels are used to guide other ants in subsequent iterations. ACO has been modified to solve a diverse class of NP-hard problems, such as finding shortest paths in packet switched networks \cite{ahn, antnet, mobile} and finding best routes in vehicle routing problems, as done in \cite{acr} and \cite{bell}.

\begin{figure}
    \centering
    \includegraphics[scale=0.7]{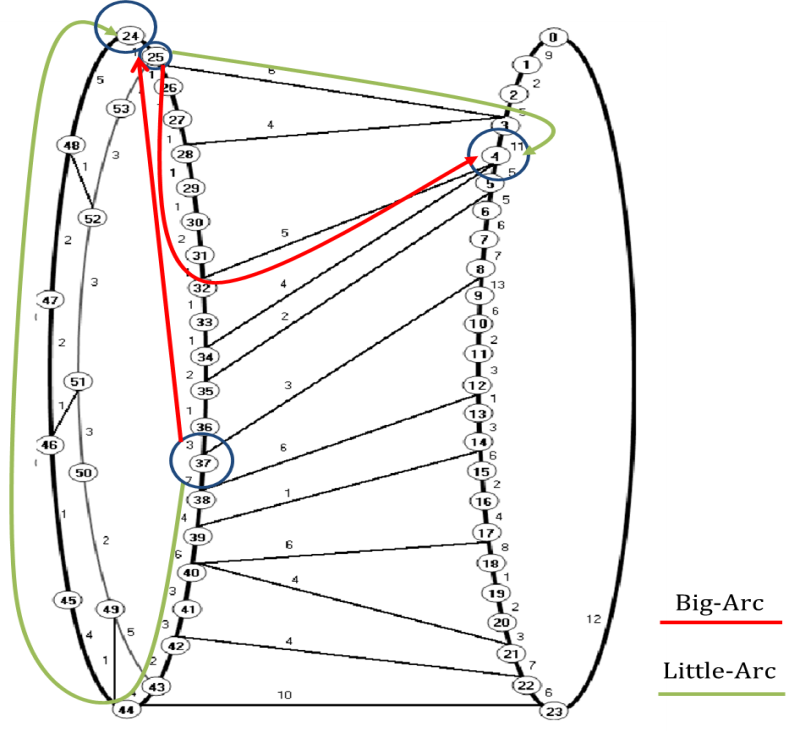}
    \caption{Two-Ring Network showing sample routing paths \cite{africa2}}
    \label{figA}
\end{figure}

\begin{figure}
    \centering
    \includegraphics{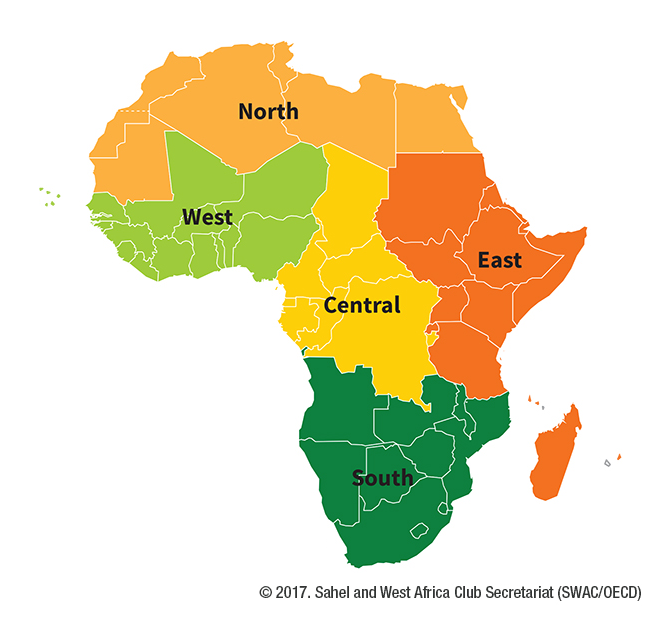}
    \caption{African countries grouped into regions \cite{SWAC}}
    \label{figB}
\end{figure}

The relatively limited number of research works directly focusing on network routing across Africa is an indication of a research gap and serves as motivation for this paper. In this work, we propose a continental network for intercommunication across African nations and use ACO for optimal path selection across these nations. Such a network can serve as a backbone for numerous collaborative health and educational frameworks being proposed for Africa. Our proposed model, called Africa 3, groups African 
countries into clusters, similar to that of the African Union (Fig. \ref{figB}) and then applies ACO to identify the most efficient network path within and across clusters. In comparing our model to the dual-ring model proposed in \cite{africa2}, rather than 
connecting individual countries directly to the dual-ring, our model connects clusters of countries together. The benefits of this approach are as follows: 
\begin{itemize}

\item{\textbf{Avoidance of single point of failure:}} with the ring topology, every node is potentially a point of failure. This is of particular concern in many African countries, where stable and uninterrupted electricity supply is still a challenge. This exponentially increases the probability of nodal failures. Though network traffic can flow in both directions in such dual-ring topology, it is not uncommon for multiple countries in Africa to experience total power failure.
 
\item{\textbf{Reduced routing distance:}} in \cite{africa2}, the Little-Arc-First (LAF), shown in Fig. \ref{figA}, was the primary routing path as it represented the shortest network distance. This might not always be the best option, as is the case of transmission between Dakar (node \#25) and Abuja (node \#4), where the LAF is the path: \#25, \#3, \#4 and costs 6 + 11 = 17. The Big-Arc on the other hand goes through the path: \#25, \#26, \#27... \#32, \#4 and has a cost of only 13. Similarly, when routing from Luanda (\#37) to Nouakchott (\#24), the LAF is the path through \#38, \#39 … \#48, \#24 at a cost of 43, while the Big-Arc through path \#36, \#35 … \#25, \#24 only costs 17. 
Using the ACO, our model ensures that the path selected between countries is always the best (lowest route cost).
\end{itemize}

\subsection{The African Continental Network Model (Africa3)}

In this subsection we present our model for interconnecting Africa. As stated earlier. it is a two level model with levels described as follows:

\textbf{Lower Level}
We begin by grouping countries in Africa into clusters, using a combination of geographical coordinates (latitude and longitude), population size, number of data centres and number of independent intercontinental submarine cable landings present in each country. These features were chosen because: i. geographical coordinates are one of the natural ways of clustering countries; ii. population size and socio-economic power influence the demand for network traffic; iii. data centre count gives a quick insight to the level of technological development, with the world rapidly adopting numerous technologies that are dependent on Cloud computing \cite{mell}; iv. submarine cable landings are connection points that link Africa to the rest of the world. These points therefore serve a vital role of gateways in our model. 

Upon successfully clustering the continent, we apply a variant of ACO called Ant-Colony Routing with Stench Pheromones (ACO-SP) to find optimal network paths within clusters and to gateways.

\textbf{Upper Level}
At this level, our model focuses on interconnecting the islands of clusters formed at the lower level. The ACO-SP is also used to find the shortest path between clusters. Of particular importance is the fact that, while seeking for optimal routes, our model strives to avoid paths that directly cross the Sahara desert. 

\subsection{Architectural Framework}

\begin{figure*}
    \centering
    \includegraphics[width=\textwidth, height = 6cm, keepaspectratio]{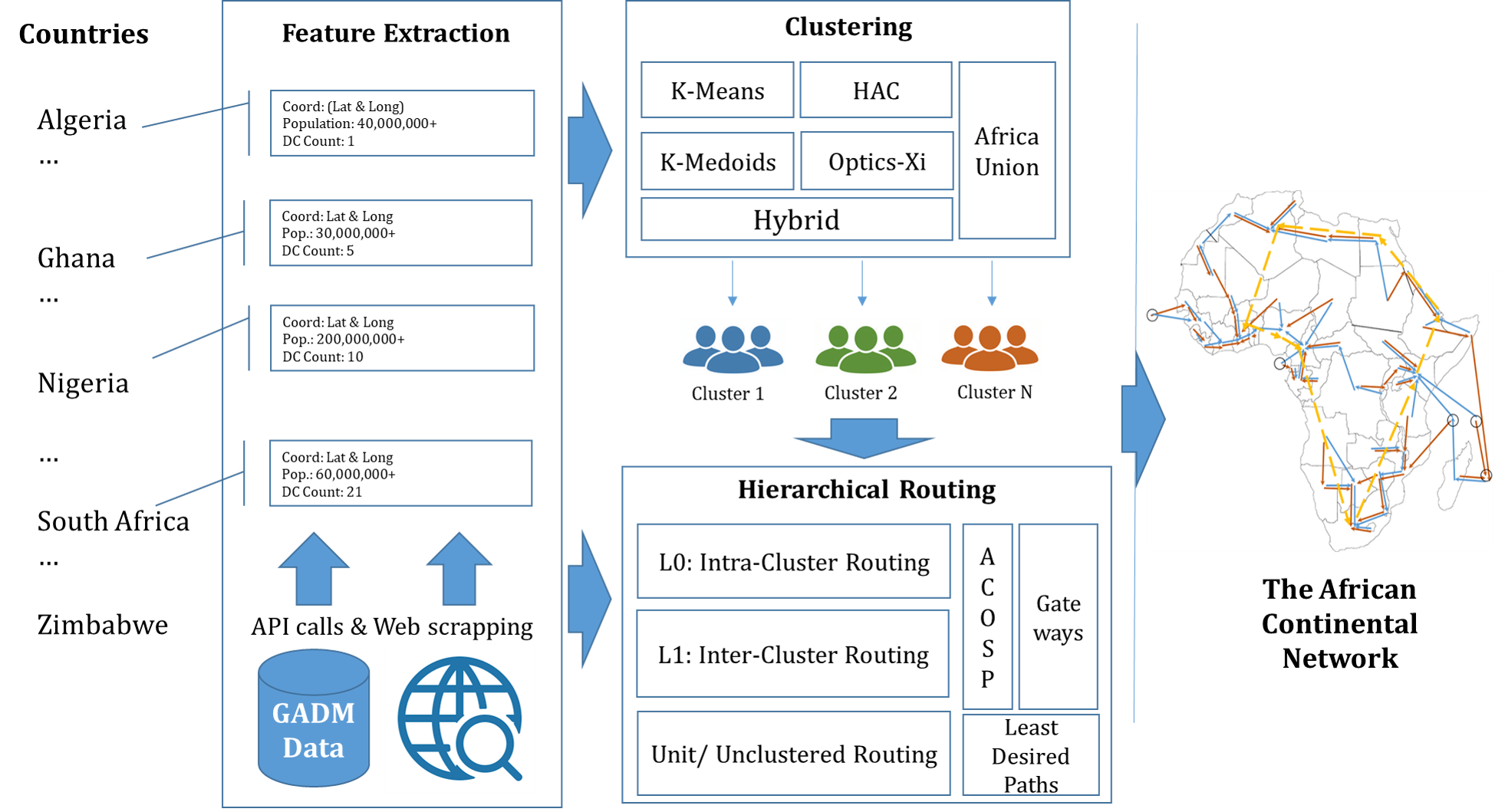}
    \caption{The African Continental Network Framework}
    \label{figC}
\end{figure*}

Fig. \ref{figC} gives an illustrative overview of our proposed continental network model for Africa. At the first phase, information about the 55 African countries are obtained. These information include: coordinates of each country's centroid (geographical mid-point), population size and number of data centres present. They were obtained by web scraping and/or API calls on \cite{dcmap, afPop, GADM}. 
These information are then fed into different clustering models as features. Here the countries are grouped into clusters, with the Africa Union's clustering used as a benchmark. This clustering phase is discussed in section \ref{s3}.

Upon successfully clustering the continent, the various clusters are fed into a routing model, where optimal paths interconnecting countries within each clusters (intra-cluster) and across clusters (inter-cluster) are obtained. The routing model is based on ACO meta-heuristic, and is discussed in section \ref{s4}. 
The final output is the African Continental Network.

\section{Africa In Clusters}\label{s3}
In this section, we focus on the lower level of our Africa 3 model and consider various algorithms for effectively splitting African countries into clusters. The African Union has grouped the continent into 5 regions, namely – Central, Eastern, Northern, Southern and Western, with member countries listed on Table \ref{tab1} and shown in Fig. \ref{figB}. In grouping countries into clusters, we considered a number of unsupervised machine learning algorithms and use the AU’s clustering layout as a comparative reference.

Table \ref{tab1} lists our input data and consists of two types of data, viz.: geo-spatial data (latitude and longitude) and temporal data (population size and data centre counts). The final feature of inter-continental submarine landings would be discussed later. In clustering these data, we considered two types of K-means clustering algorithms, a hierarchical clustering algorithm and a density-based clustering algorithm. These algorithms are described as follows:

\begin{itemize}
    
\item{\textbf{Classic K-means (Lloyd-K-Means):}}
The classic K-Means is a centroid based clustering algorithm, which, in essence, implies that it determines clusters members by how close data points are to a centre point \cite{km}. For k-means, we initially set the value of k to 5, in line with the AU’s regional distribution of the continent into five regions.  Rather than solely relying on the AU’s clustering for the value of k, we also used two elbow methods - the Distortion score and Calinski-Harabasz score \cite{sklearn} to determine the optimal value of k. We then repeated the clustering using this new optimal value of k. Euclidean distance \cite{euclidean} was used as proximity metric for selecting cluster members. Our decision to use the Euclidean distance was because Africa is uniquely located almost symmetrically around the equator (latitude $0^o$) and the prime meridian (longitude $0^o$), at $30^o W, 60^oE, 40^oN, 40^oS$; the use of the Euclidean distance is therefore justifiable as there are no longitudinal wrap around ($0^o$ and $180^o$). Despite this justification and a similar other \cite{Montreal}, for thoroughness and avoidance of doubt, we also clustered using the Great Circle Distance (Haversine) \cite{haver}. We implemented this with the k-Medoids algorithm described next.

\item{\textbf{K-Medoids:}} K-Medoids or Partitioning Around Medoids (PAM) algorithm, like K-means is also a partitioning algorithm which attempts to group similar data together based on distance. However, unlike k-means the cluster centre in K-Medoids is one of the data points rather than an average of the points in the cluster as is the case with k-means. K-Medoids also allows the use of arbitrary distance metric \cite{kmedoid}. With the earth being spherical, it has been argued that the Euclidean distance might be less than ideal for clustering geo-spatial data. To this end, we implemented the K-Medoids algorithm with the Haversine distance metric \cite{haver}, described in \eqref{eq1}.

\begin{equation}
\begin{array} {lr}
    distance = R * c    \\
    c = 2 * atan2 (\sqrt{a}, \sqrt{1-a})     \\
    
    a = \sin^{2}(\frac{(lat_2 - lat_1)}{2}) + \\
    \cos{(lat_1)} * \cos{(lat_2)} * \sin^{2}(\frac{(lon_2 - lon_1)}{2}) 
    
    \end{array}
    \label{eq1}
\end{equation}

where $lat_{i}$ and $lon_{i}$ are the latitude and longitude of the respective countries $\imath$ in Radians and $R$ is the radius of the earth.

\item{\textbf{Hierarchical Agglomerative Clustering (HAC):}} This is a hierarchical “bottom-up” clustering technique that treats each data point (country) as a unit cluster and pairs up similar clusters. This process is repeated until all nodes are grouped into the minimum number of clusters \cite{hac}. We used two pairing metrics – the Euclidean distance and Complete Affinity. We chose the Complete Affinity because we sought to maximize the distance between data points. The advantage of this becomes more apparent when we cluster based on the geo-spatial, population and data centre sizes; as having multiple countries with large population sizes within the same cluster is undesirable. Similarly, for data centres, it is also desirable to have widely (and evenly) dispersed data centres across the continent, in order to improve the overall technological capabilities of the continent as a whole.

\item{\textbf{OPTICS-Xi Clustering:}} The Ordering Points To Identify the Clustering Structure-Xi (OPTICS-Xi) algorithm \cite{optix} is based on the Density-Based Spatial Clustering of Applications with Noise (DBSCAN) clustering algorithm, which has been reported to be better for clustering geo-spatial data. OPTICS-Xi is a variant that relaxes the need to set correct values of epsilon $(\epsilon)$, which is often difficult to determine for most data sets \cite{optix, rdrr}. For our implementation, we considered a country as a point $P$ and then used the OPTICS-Xi clustering to find the set of countries that were within a “circular radius” $(\epsilon)$ of $P$. Unlike the DBSCAN, OPTICS-Xi relaxes the need to set a value for epsilon. However, we set minPts (neighbourhood density) to 3, thereby ensuring that each cluster had at least 3 countries. 

\item{\textbf{Multi-Feature Clustering:}}
We combined all features from our dataset and cluster based on this. Two clustering techniques were used here: i. a hybrid of K-Medoids and HAC; ii. OPTICS-Xi. For this combined feature clustering we replace both Euclidean and Haversine distances with the weighted distance defined in \eqref{eq2}.
 
 \begin{equation}
 \begin{array} {lr}
weightedDistance = \alpha  * hdistance + \\
\beta * population + \gamma * dcsize    
\end{array}
\label{eq2}
 \end{equation}
	
where \textit{hdistance} is the haversine distance between two countries, \textit{population} is the population size and \textit{dcsize} is the number of data centres present. $\alpha, \beta, \gamma$ are weight factors, which respectively determine the influence of \textit{hdistance, population and dcsize}. 

With these 3 features (co-ordinates, population and dcsize) being of different scales, normalization was required. For normalization we used the min-max normalization \cite{nom1, nom2} defined in \eqref{eq3}, with resulting values scaled between $0$ (relative min. value) and $1$ (relative max. value).

\begin{equation}
    X’ = \frac{x - min(x)}{(max(x)- min(x))} 
    \label{eq3}
\end{equation}

Once the clusters have been obtained, we then ACO-SP for path finding. The processes involved are described in the next section.

\end{itemize}

\section{Continental Network Grid} \label{s4}
One of the primary objectives of this work is to propose a network to interlink countries in Africa for the purpose of collaborating and sharing network infrastructure. Recent evidence has shown an imbalance in the level of development of Africa countries. Some countries are comparatively at the forefront of innovation and technological advancements, while others are stuck in the middle age. By leveraging on collaborative technologies, growth and advancements in many developing countries of Africa can be accelerated. Such collaboration could enhance growth in various areas including but not limited to health care, trans-boundary water management and general socio-economic activities. However, in order to achieve this, an underlining and robust network needs to be put in place. 

\subsection{Selecting Continental Gateways}

We define continental gateways as landing points for trans-continental submarine fibre optics cables, which provide Internet connectivity to the continents. There are a number of continental gateways across Africa, most of which are located in coastal cities. Table \ref{tab2} gives a concise summary of these continental gateways in Africa \cite{sub}. The table shows that there are about 13 cities with trans-continental submarine fibre optics cables terminals in Africa.  Note that, though EASSy has been included, it is not a trans-continental fibre optic line as it does not link Africa to any other continent. These continental gateways would serve as end-point for our continental network grid. There are a number of them across Africa, among which we choose a few, which we call Prime Continental Gateways (PCGs). For this work, a PCG is defined as any African country having multiple independent submarine cables landing within her shores.  Using Table \ref{tab2} as a guide, we chose countries with 5 or more trans-continental submarine fibre optics cables landings. These include: Algeria (5), Nigeria (8), Cameroon (6), South Africa (8), Kenya (6), Djibouti (9), Egypt (15), and Libya (5).  Though Cameroon and Nigeria are in close proximity, the huge population (and by extension network demand) of Nigeria justifies the selection of Cameroon as a PCG. 

\begin{table}
\caption{Major Trans-continental Submarine Fibre Optics Landings in Africa.}
\label{tab2}
\begin{scriptsize}
\begin{tabular}{ p{0.3cm}|p{1.5cm}|p{2cm}|p{1.5cm}|p{4cm}|p{0.25cm}}
\hline
\textbf{S/N} &	\textbf{Name} & \textbf{Source} & \textbf{Destination} & \textbf{Counties Traversed} & \textbf{Country Count}
\\ \hline     
1 & MainOne & Portugal & Nigeria (Lagos) &	Ghana & 	3     \\ \hline
2 & SAT3 / WASC &	Spain & South Africa (Western Cape)	& Senegal, Côte d'Ivoire, Ghana, Benin, Nigeria, Cameroon, Gabon and Angola	& 12 \\ \hline
3 & South Africa Far East (SAFE) &	South Africa (Western Cape)	& Malaysia (Penang) &	Africa (Mauritius), Réunion and India 	& 5 \\ \hline
4 &	West Africa Cable System (WACS) & 	United Kingdom	& South Africa (Western Cape) &	Africa (Namibia, Angola, DRC, Congo, Cameroon, Nigeria, Togo, Ghana, Côte d'Ivoire, Cape Verde, Canary Islands), Portugal 	& 14 \\ \hline

5 & Seacom	& South Africa (Cape Town) &	India (Mumbai) 	& Africa (Tanzania, Djibouti, Kenya Mozambique, Egypt, Botswana, Uganda, Rwanda), Saudi Arabia, India, France, Germany, Netherland, UK.	& 16 \\ \hline
6 &	EASSy &	South Africa (KwaZulu-Natal) & Sudan &	Tanzania, Djibouti, Mozambique, Somalia, Kenya, Comoros, Madagascar	& 9 \\ \hline
7 &	The East African Marine System (TEAMS) &	Kenya (Mombasa) &	UAE (Fujairah)	& &	2 \\ \hline

8 &	South Atlantic Inter Link (SAIL) &	Brazil (Fortaleza) & Cameroon (Kribi)	& &	2 \\ \hline

9 &	South Atlantic Cable System (SACS) &	Brazil (Fortaleza) & Angola (Sangano) & &	2 \\ \hline

10 &	Asia Africa Europe (AAE-1) &	China (Cape D’Aguilar) &	France (Merseille) &	Vietnam, India, Cambodia, Oman, Thailand, UAE, Myanmar, Qatar, Malaysia, Italy, Pakistan,Yemen, Saudi Arabia, Greece, Africa (Egypt, Djibouti) & 18 \\ \hline

11	& Europe India Gateway (EIG) &	India (Mumbai) &	UK (Bude) &	Oman, UAE, Saudi Arabia, Gibraltar, Monaco, Portugal, Africa (Egypt, Libya, Djibouti)	& 12 \\ \hline

12 & SeaMeWe-3 &	Australia &	Europe &	Numerous Asian and European Countries + Africa (Egypt, Djibouti) & 28 \\ \hline
13 &	SeaMeWe-4, SeaMeWe-5+ &	France (Toulon) &	Malaysia (Melaka) &	European, Euro-Asian countries, Africa (Algeria+, Tunisia+, Egypt and Djibouti) &	15, 17+ \\ \hline

14	& FALCON &	Egypt (Suez) &	India (Mumbai) &	Iraq, Yemen, Saudi Arabia, Kuwait, Oman, Iran, Qatar, UAE, Bahrain, Africa (Sudan)	& 12 \\ \hline

15	& PEACE &	Kenya (Mombasa)	& France (Marseille)	& Africa (Egypt, Somalia, Djibouti, Seychelles), Pakistan &	7 \\

  \hline
\end{tabular}
\end{scriptsize}

\end{table} 

\subsection{Traffic Routing}
We apply a modified ACO to find optimal network paths across the continent. This ACO is applied twice in our continental model; first at the cluster level and then at the continental level. At the lower hierarchy (cluster level), ACO is used to find paths from cluster members to the cluster head within a cluster. At the continental level, it is used to find ideal paths between the various cluster heads and the PCGs. For comparative purposes, we also considered an ‘unclustered’ approach, wherein ACO is applied directly at the continental level. In this model, all $55$ member countries of Africa (54 + Western Sahara), are taken as nodes and ACO is applied to obtain an optimal network paths across the entire continent and to the PCGs.

In developing this continent-wide network for Africa, we considered a variant of ACO called Ant Colony Optimization with Stench Pheromone (ACO-SP) \cite{aco-sp}. ACO-SP has been used in network packet routing with the primary objective of preventing congestion. In a typical ACO algorithm, the ants obtain the best route (often the shortest route) across a network. The problem with this, in a packet routing network, is that all traffic would flow through this “best route”, thereby: i. under-utilizing other routes, ii. resulting in network congestion on the optimal route. ACO-SP, therefore, introduces a second pheromone called “stench pheromone” used to discourage selection of a given path, thus pushing some ants away from the best path and forcing them to create alternate best path(s). 

Applying this concept, we created a list of undesirable paths, which includes paths that directly traverse the Sahara desert. Conversely, we also created a list of desired paths, which are paths we encourage the ants to traverse and consists of the PCGs. In implementing this, we created a multiplicative index, which either increases the rate of pheromone evaporation for undesired paths or reduces it for desired paths.

Our implementation process is as shown in Algorithm \ref{alg1} with parameters detailed on Table \ref{tab3}. On the table, \textit{f} is used by the ACO-SP algorithm to introduce a level of randomness into the system. \textit{s} and \textit{V} are variables which indicate the location of ants; their values ranges between 1 and 55, with each location representing an African country. 1000 ants were used in the experiment and this is indicated with variable \textit{k}. $\Delta$ is a variable representing the weighted distance between two countries and is defined by \ref{eq2}. $\rho$ is a factor which determines how fast the pheromones evaporate. It is a unique  feature that distinguishes our work from \cite{aco-sp} and is defined in step 5, equation \ref{eq8} of Algorithm 1. Finally, S and U are sets of countries with PCGs and within the Sahara desert respectively.   

\begin{table}
\caption{Parameters and Settings}
\label{tab3}
\setlength{\tabcolsep}{3pt}
\begin{tabular}{c c l }
\hline
\textbf{Variable}   &  \textbf{Value} & \textbf{Description} \\ \hline
$f$ & $0-1$ & 	Used to introduce randomness into the system \\ \hline
$T$ & $0.8$ & Threshold value \\ \hline
$s$ &$1-55$& Current location of ant \\ \hline
$d$ &$1-55$& Next location of ant \\ \hline
$k$ & $1-1000$	& Individual Ant  \\ \hline
$V$ &	$1-55$ & List of countries visited by ant k \\ \hline
$\Phi$ & $0.2$ & Initial Pheromone level \\ \hline
$\alpha$ & $0.33$ &	Determines the influence of distance \\ \hline
$\beta$ & $0.33$ &	Determines the influence of population \\ \hline
$\gamma$ & $0.33$ &	Determines the influence of DC count \\ \hline
$H_{sd}$	& Equation \eqref{eq1} &	Haversine distance between two countries s and d \\ \hline
$Q$ & &	Population of destination country \\ \hline
$C$ & &	Data centre counts of destination country \\ \hline
$\Delta$ & Equation \eqref{eq2} &	Route cost (weighted distance) between two countries.  \\ \hline
$\rho$ & $0.2$	& Pheromone evaporation rate \\ \hline
$P_{sd}$ & &	Probability of moving from country s to country d \\ \hline
$G$ &	[PCGs] & Set of desirable paths \\ \hline
$U$ & &		Set of undesired paths \\ \hline
\end{tabular}
\end{table}

\begin{algorithm}
1. Each ant starts from a randomly selected country.\;

2. Generate a random number f, f = [0, 1] and a threshold value (0.8) \;

3. From any given country ($s$), the ants choose the next country ($d$) to traverse to using \eqref{eq4} adapted from \cite{bell}. This strives to find the maximum ratio between pheromone level and distance as follows \;
\\
If {$f\leq threshold$} 
        \begin{equation}
        d^k = max(\frac{\Phi_{sd}}{\Delta_{sd}}), d^k \not \in V^k    
        \label{eq4}
        \end{equation}
        
In calculating our route cost $\Delta$, we used a weighted distance metric based on \eqref{eq5}, where $\alpha + \beta + \gamma = 1$.
\;       

\begin{equation}
\Delta_{sd} = \alpha * H_{sd} + \beta* Q_{d} + \gamma * C_{d}			
\label{eq5}
\end{equation}

If {$f > threshold$} 
    \\
    The next country is calculated based on a probabilistic function which favours the shortest path with the highest pheromone level. This probability is given in \eqref{eq6} described as follows: \;

    \begin{equation}
        P_{sd} = \frac{\Phi_{sd}} {\Delta_{sd}* \sum_{su} \frac{\Phi_{su}}{\Delta_{su}}}, u,d \not \in V^{k}
        \label{eq6}
    \end{equation}

4. Update Pheromone level based on \eqref{eq7} described as: \;

\begin{equation}
    \Phi_{sd} = (1- \rho_{sd}) * \Phi_{sd} + \frac{\rho_{sd}}{\Delta}
    \label{eq7}
\end{equation}

5. Adjust evaporation rate: in our case evaporation rate is not constant across the entire system but rather varies for specific trails to encourage or discourage the use of certain trails. Evaporation rate is thus adjusted based on the conditions given in \eqref{eq8}.
\;

\begin{equation}
\rho_{sd}=
\left \{
\begin{array} {lr}
\rho, d \not \in G,U \\
\rho*0.25,d \in G   \\
\rho*0.75, d \in U) \\
\end{array}\right.
    \label{eq8}
\end{equation}

$G$ and $U$ are respectively sets of encouraged trails (such as PCGs) and set of discouraged trails (trails through deserts). \;

 \caption{Modified Ant Colony Routing with Stench Pheromones}
 \label{alg1}
\end{algorithm}

\section{Simulations}\label{s5}
Simulation were carried out on a system configured with Intel core i7 processor, 8GB RAM and Windows 10 Operating System. Python (Jupyter notebook with Scikit Learn library \cite{scikit}) and Java (ELKI Data mining framework for Java \cite{elki} with Eclipse IDE) were used, as both offered complementary libraries and visualization tools.

Geo-spatial data of each country were scrapped from \cite{GADM}, while data on population were obtained from  \cite{afPop} and DC size from \cite{dcmap}. After normalizing (\ref{eq3}), the data were passed into the respective clustering module. K-Means, HAC and Optic-Xi clustering were done in Python, while K-Medoids and the combined clustering were done using ELKI. This was because ELKI had in-built support for K-Medoids and also provided the ability to use custom distance function (such as \ref{eq2}), which was vital for this work. However, a major disadvantage of ELKI was that, its maps were in gray scale and it only gave cluster points. We therefore had to manually colour code each of the resulting clusters, and superimpose the outline of the map of Africa.

\subsection{Clustering}
As described above, our Africa 3 model groups African countries into clusters. For clustering, we compared four different clustering algorithms namely K-Means, KMedoids, OPTICS-Xi and HAC; using two distance metrics - Euclidean distance and great circle distance (Haversine). We carried out our clustering in multiple phases. At the first phase, we clustered using only geometric values (i.e. Latitude and Longitude). Results of this phase are shown in Fig. \ref{figW} to \ref{figYb}. We then clustered based on DC size and populations. Being 1 dimensional, the results were simple histograms of country versus frequency (count). These are not shown for purpose of space savings.  Afterwards, the three features were combined for clustering and obtained results are shown in Fig. \ref{figZa}, \ref{figZb} and \ref{figZc}. The AU clustering map, has been included in each result for reference purposes. 

\subsubsection{Clustering using K-Means}

\begin{figure}
\centering
\subfigure[Calinski Harabasz Elbow Score]{
\includegraphics[scale=0.3]{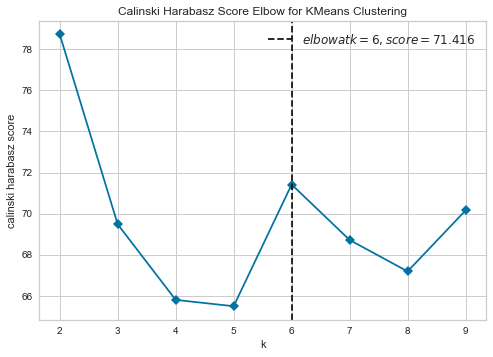}
\label{figR1a}
}
\subfigure[Distortion Elbow Score]{
\includegraphics[scale=0.3]{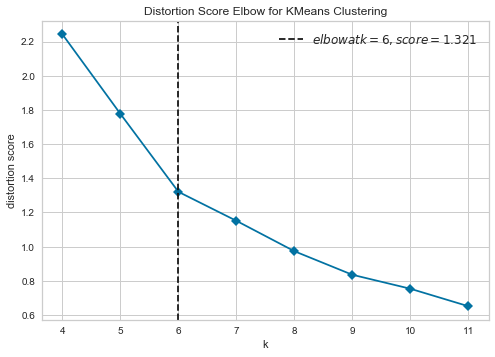} 
\label{figR1b}
}
\caption{Selecting ideal value of K} \label{figR1}
\end{figure}

\begin{figure}
\centering
\subfigure[K-Means clustering with k=5]{
\includegraphics[scale=0.50]{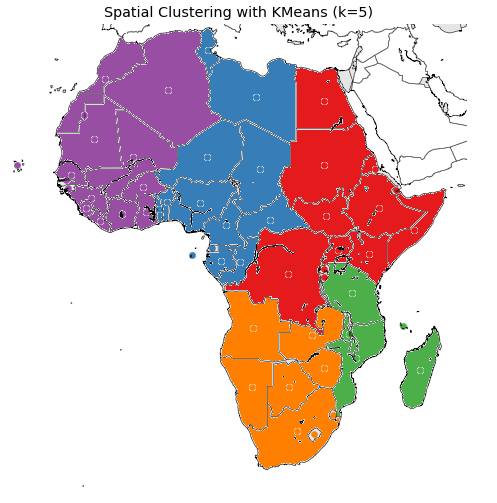} 
\label{figWa}
}
\subfigure[K-Means clustering with k=6]{
\includegraphics[scale=0.5]{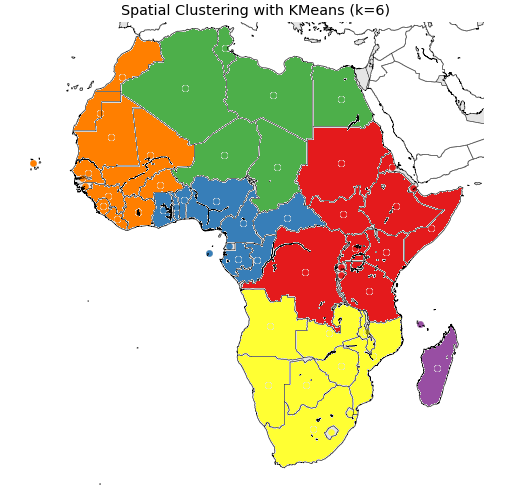} 
\label{figWb}
}
\subfigure[AU’s clustering]{
\includegraphics[scale=0.5]{images/AUafrica.png} 
\label{figWc}
}
\caption{K-Means Clustering using Euclidean Distance}\label{figW}
\end{figure}

Fig. \ref{figWa} shows clustering results when K-Means is used, with the value of $k$ set to 5 and the Euclidean distance used. The choice of $k=5$ was done purely for the purpose of comparing our results with AU’s which also has 5 clusters. In Fig. \ref{figWb}, rather than fully relying on the AU’s classification and setting $k$ to 5, we instead chose the value for $k$ by running two elbow methods - the Distortion score and Calinski-Harabasz score \cite{sklearn}. Both of them gave a value of 6 for $k$, as shown in Fig. \ref{figR1a} and \ref{figR1b}. For both values of $k$ (5 and 6), the resulting cluster maps differ from that of the AU’s. This is most prominent in the Northern and Western clusters.

\subsubsection{Clustering using K-Medoids}
\begin{figure}
\centering
\subfigure[K-Medoids clustering with k=5]{
\includegraphics[scale=0.50]{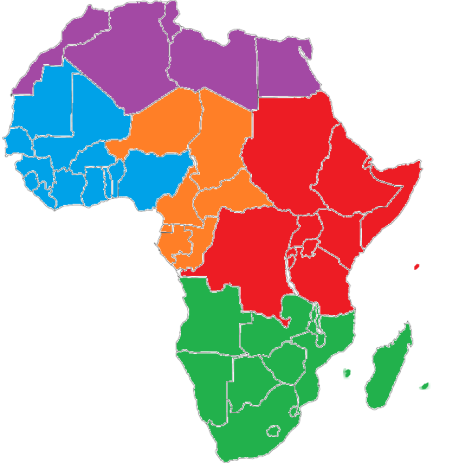} 
\label{figXa}
}
\subfigure[K-Medoids clustering with k=6]{
\includegraphics[scale=0.50]{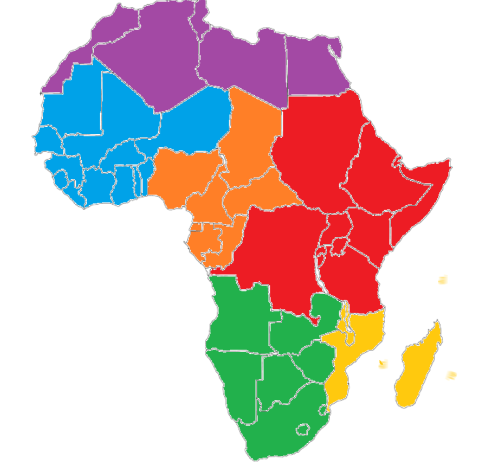} 
\label{figXb}
}
\caption{K-Medoids Clustering using Haversine Distance}\label{figX}
\end{figure}

We then proceeded to cluster based on the great circle distance, a metric which takes the shape of the earth into consideration, when measuring the distance between points. However, it has been reported that K-Means is not well suited for use with distance metric other than Euclidean, for this purpose we also clustered using K-Medoids. Obtained results are depicted in Fig. \ref{figXa} and \ref{figXb}, for values of $k= 5$ and $6$ respectively. 

For both values of $k$, obtained results were similar to the AU’s clustering. When $k$ was set to 5, Fig. \ref{figXa} shows Madagascar, Comoros and Mauritius being grouped in Southern Africa. The DRC was classified as part of Eastern Africa, while Niger was removed from Western Africa and included in Central Africa. Finally Mauritania was classified as a West African country. 

With $k$ set to 6, similar results were obtained for Northern and Eastern Africa, with the only exception being Seychelles which was put in a separate cluster, alongside Madagascar, Comoros, Mauritius, Mozambique and Malawi. To the West, Nigeria was classified as a Central African country, while Niger took its place in West Africa. These are as shown in Fig. \ref{figXb}. 

\subsubsection{Clustering using HAC}

\begin{figure}
\centerline{\includegraphics[scale=0.8]{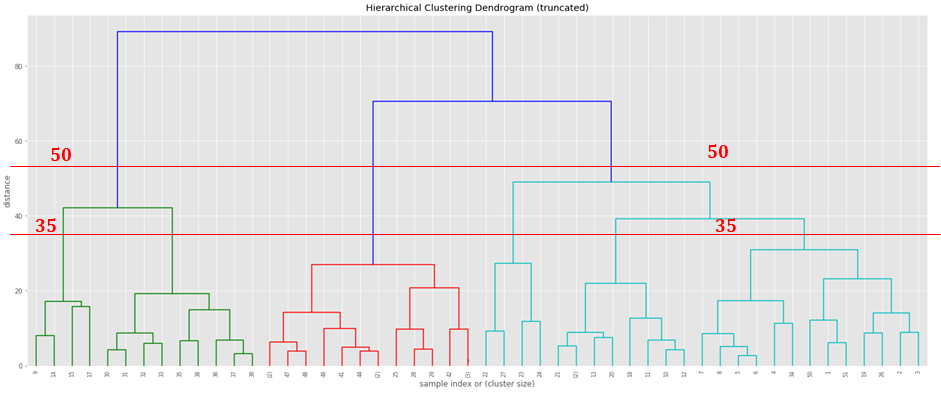}}
\caption{HAC Dendrogram (Distance = 35, 50) .\label{figYa}}
\end{figure}

\begin{figure*}
\centerline{\includegraphics{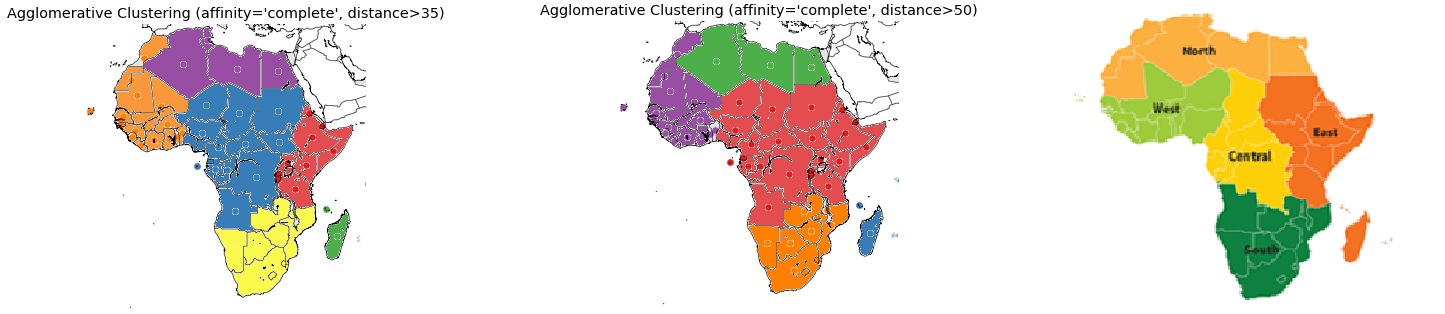}}
\caption{Hierarchical Agglomerative Clustering (HAC) – 35 and 50) .\label{figYb}}
\end{figure*}

We then considered a third clustering algorithm - the HAC, which is an Agglomerative clustering technique. The advantages of this algorithm are its efficiency with small datasets and its implicit ability to determine the ideal number of clusters. Fig. \ref{figYa} shows the Dendrogram, from which we chose distances of 35 and 50. This translated to 5 and 6 clusters, with the resulting cluster diagrams shown in Fig. \ref{figYb}. Results obtained in both cases were highly skewed towards Central Africa, with Northern and Eastern Africa having only 4 countries each. 

\subsubsection{Multi-feature clustering}
From the above results, only the K-Medoids algorithm gave results closest to the AU’s clusters; for this purpose we decided to stick with it. Though we also clustered based on population and data centre size, the obtained histograms are not shown. We instead show the result of combining all three features (geographical coordinates, population and data centre size). Due to the widely dispersed values, we normalized these data using the Min-Max normalization.

\begin{figure}[hbp!]
\centering
\subfigure[Using K-Medoids (k=5)]{
\includegraphics[scale=0.6]{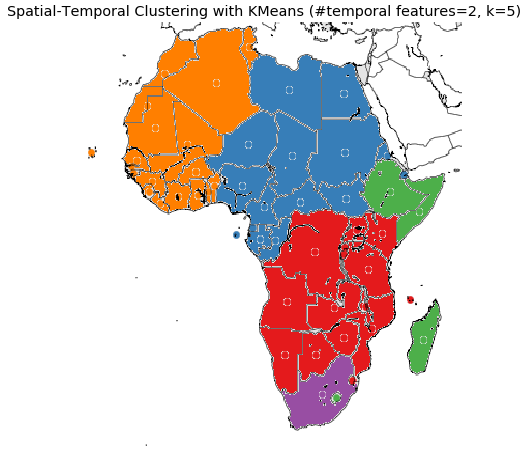} 
\label{figZa}
}
\subfigure[Using K-Medoids (k=6)]{
\includegraphics[scale=0.6]{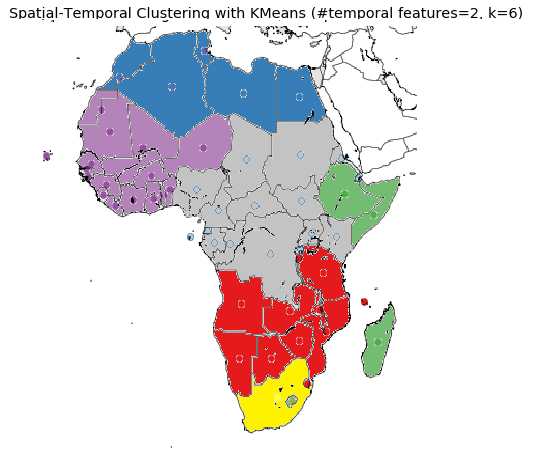} 
\label{figZb}
}
\subfigure[Using OPTICSXi (minPt=3)]{
\includegraphics[scale=0.6]{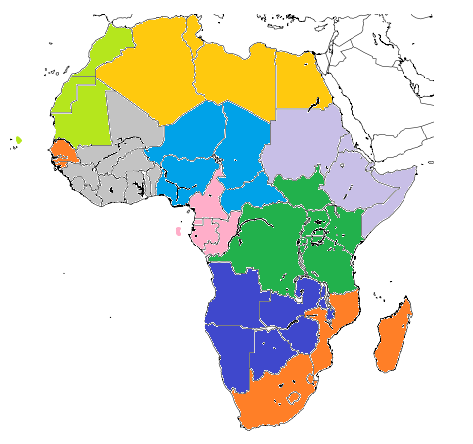} 
\label{figZc}
}
\caption{Combined Feature Clustering}\label{figZ}
\end{figure}

\begin{figure}
\centerline{\includegraphics[scale=1]{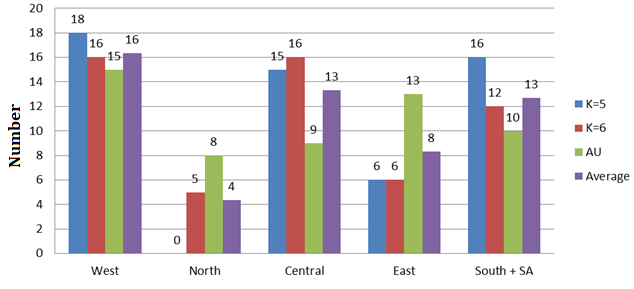}}
\caption{Comparison of Country Count per Cluster  .\label{figM}}
\end{figure}

We compared the K-Medoids algorithm (for $k$ values of 5 and 6) based on the combined featured, and using Haversine distance as metric. Fig. \ref{figZa}, \ref{figZb}, \ref{figZc} show the obtained clustering results. For comparative purposes, we also ran the OPTICSXi algorithm (which is a density based clustering technique) using these combined features. In Fig. \ref{figZa}, countries from the original AU’s clustering were remapped into entirely new clusters. The West was made up of countries drawn from both Northern and Western African countries of AU’s clustering. Similar reconfiguration were observed in the Central African cluster. These can be attributed to the influence of the geo-spatial features of the data set. To the South, clustering similar to that of the AU was observed; however, South Africa was singled out as a unit cluster. This is certainly due to the influence of data centre size feature, as South Africa has a significantly higher number of data centres than any other country in the Southern Africa region and the entire continent as a whole. Finally, the effect of population is obvious at the Eastern Cluster, as Ethiopia (with a significantly higher population) skews the clustering in its favour, resulting in the formation of a new cluster. The influences of the individual features on the overall clustering are also visible in the other results shown in Fig. \ref{figZb} and \ref{figZc}.  A comparative distribution of countries into clusters for the K-Medoids and AU is shown in Fig. \ref{figM}.

Fig. \ref{figM} shows the distribution of countries across the three clustering layouts, viz.: multi-feature clustering with $k$ set to 5 (Fig. \ref{figZa}), combined feature clustering with $k$ set to 6 (Fig. \ref{figZb}) and African Union clustering layout (Fig. \ref{figWc}). In all three models, West Africa had the highest number of countries.  Both K-Medoids models ($k$=5 and 6), allocated 6 countries to East Africa, which is less than half of the AU’s (at 13). With the $k=5$ K-Medoids model, the North African cluster was completely removed, while $k=6$ classified 5 countries as Northern countries versus AU’s 8. K-Medoids  grouped more countries into Central and Southern Africa than the AU did. 

A summary of the clustering algorithms considered and their respective parameters is given in Table \ref{tab4}.

\begin{table}
\caption{Summary of Clustering Results}
\label{tab4}
\setlength{\tabcolsep}{3pt}
\begin{tabular}{p{1.8cm} p{2cm} p{1.5cm} p{1.5cm} p{2cm}}
\hline
\textbf{Algorithm} & \textbf{Clustering Metric} &	\textbf{Distance Metric} &	\textbf{Clusters Count} & \textbf{Attributes} \\ \hline
K-Means	& $k=5$ & Euclidean & 5 & Lat. \& Long. \\ \hline
K-Means	& $k=6$	& Euclidean & 6 & Lat. \& Long.
\\ \hline
KMedoids & $k=5$ & Haversine & 5 & Lat. \& Long.
\\ \hline
KMedoids & $k=6$ & Haversine & 6 & Lat. \& Long.
\\ \hline
HAC & $Dist.=35,50$	& Complete Affinity &	6 &	Lat. \& Long. \\ \hline
Multi 
[K-Medoids + HAC] & $k=5$ &	Haversine &	5 &	Lat., Long., Pop. \& DC count \\ \hline
Multi 
[K-Medoids + HAC] &	$k=6$ & Haversine &	6 &	Lat., Long., Pop. \& DC count \\ \hline
Multi
[OPTICSXi] & $points = 3$ & Haversine & 10 &	Lat., Long., Pop. \& DC count \\ \hline
Africa Union &	NA	& NA &	5 &	Lat., Long. \& Geo-political \\ \hline

\end{tabular}
\end{table}

\subsection{Routing}

Having put the countries into respective clusters, we now proceed with creating network(s) to interconnect countries across the continent. Our proposed model first creates an inner-network to link countries within clusters (intra-cluster routing); then creates a second high level network between clusters and to PCGs (inter-cluster routing). For comparison purposes, we also created a third network that treats each country as a cluster in itself and interconnects all unit countries to themselves. The Total Route Cost (TRC) is used as a metric to measure how good the selected route is. It is a composite metric derived from combining the population size, data centre count and geographical distance. Lower TRCs are desirable.  

\subsection{Intra-Cluster Routing}

In this section, we present intra-cluster routing result for the various clustering layout obtained in section \ref{s5}-A. Obtained results give insights into the best network route from each African country to its closest continental gateway and how Africa as a continent can be interconnected. For each experiment 1000 ants were used. The cost metric used is a value representing the shortest path cost for a network which interconnects a country to its closest continental gateway (PCG) within each cluster. Less Desirable Trails (LDTs) are network paths which cut across the Sahara desert. Our model sought to discourage the use of these paths.For each country, unique network paths were obtained and are  summarized on Tables \ref{tab6}, \ref{tab7}, \ref{tab8} under the appendix section. 

\subsubsection{Network Paths for $k=5$ clusters}
The combined feature clustering with $k=5$, resulted in five clusters and is depicted in Fig. \ref{figZa}. Our proposed modified ACO-SP algorithm was used to obtain optimal paths between each country and the respective PCG for each of  the clusters. The obtained simulation results are summarized on Table \ref{tab6}.

On the table, four clusters are presented instead of five. This is because rather than having South Africa as a unit cluster, we added it to the Southern African cluster. For the Western cluster, since none of our selected PCGs fall into the cluster, we set Ghana as the gateway. This was because Ghana (3) had the highest number of independent submarine landings of all the countries in the cluster (see Table \ref{tab4}). By doing this, we encouraged more traffic to flow through Ghana. 

For the Central, Southern and Eastern African clusters, we set their PCGs to Cameroon, Djibouti, Nigeria; Kenya, South Africa; and Mauritius respectively. Similarly, we set Tunisia, Mali, Western Sahara and Mauritania as LDTs for the Western African cluster; Niger and Sudan as LDTs for the Central African cluster. No LDTs were set for the Southern and Eastern African cluster as there are no major deserts in these parts of Africa.

\subsubsection{Network Paths for $k=6$ clusters}

We repeated the experiment for the second clustering layout shown in Fig. \ref{figZb}, where $k$ was set to 6. Obtained results are as summarized on Table \ref{tab7} (appendix section). Similar to the previous results, South Africa was included as part of the Southern cluster, hence five clusters are shown instead of six. These are Western, Northern, Central, Southern and Eastern Africa clusters. 

For the Western cluster, Ghana was again set as the gateway; while Mali, Western Sahara and Mauritania were set as LDTs. For the Northern African cluster, Algeria and Tunisia were set as the PCG and LDT respectively. For the Central African cluster, the PCGs were Cameroon, Djibouti, Kenya, Nigeria; while the Chad and Sudan were set as LDTs. For the Southern African cluster, South Africa was set as the PCG; while Mauritius was set as the PCG for the Eastern African cluster. For both the Southern and Eastern African clusters, no LDTs were set. The Table shows the optimal network path for each African country when $k=6$ model was used.

\subsubsection{Network Paths for AU clusters}

For purpose of comparison, we also repeated the experiment to obtain the routing path for each country using the AU’s clustering layout. Obtained results are as summarized on Table \ref{tab8} (appendix section). Five clusters namely Western, Northern, Central, Southern and Eastern Africa Clusters were presented. 

Nigeria was set as the PCG for the Western cluster, while Mali and Niger were the LDTs. For the Northern African cluster, Algeria was the PCG used, while the LDTs consisted of Mauritania, Sudan, Tunisia, Western Sahara, being countries in/around the Sahara Desert. For the Central African cluster, Cameroon and Chad were set as the PCG and LDT respectively. For the Southern African cluster, South Africa was set as the PCG, while Kenya and Djibouti were the PCGs for the Eastern African cluster. For both the Southern and Eastern African Clusters, no LDTs were set. The Table shows the optimal network path for each country when Africa is split into clusters based on the AU's regional model.

\subsection{Comparing Optimal Paths}
For each cluster, we also performed experiments to determine the "optimal path". For this work, an optimal path refers to a unique route, that traverses through all the countries within the cluster only once. This metric is useful in determining the best way to interconnect all countries within a cluster, while utilising the least amount of resources (e.g. Fibre optic or Ethernet cable). The result for each of the clusters within the three clustering scheme considered ($k=5$, $k=6$ and AU) are as follows:

\begin{itemize}
\item {$k=5$\bf{:Western Africa Cluster.}}
We used the following experimental parameters:
PCG: [Ghana]. 
LDTs: [Tunisia, Mali, Western Sahara and Mauritania].
Optimal network route: [Morocco, Algeria, Mauritania, Tunisia, Western Sahara, Cabo Verde, Senegal, Mali, Burkina Faso, Benin, Togo, Ghana, Gambia, Guinea Bissau, Liberia, Sierra Leone, Guinea, Cote d'Ivoire].
Total Route Cost (TRC): 43.

\item $k=5$\textbf{:Central Africa Cluster.}
PCGs: [Cameroon, Djibouti, Nigeria].
LDTs: [Niger and Sudan].
Though Egypt and Libya are also located in the Sahara, they are not included in the LDT nor are they included as PCGs. This is because the influence of the stench multiplier (for undesired paths) and boost multiplier (for desired paths) cancel each other out, hence no point including them in any of the lists.

Optimal network route: [Republic of the Congo, Gabonese Republic, Equatorial Guinea, Sao Tome and Principe, Cameroon, Nigeria, Chad, Niger, Sudan, Libya, Egypt, Eritrea, Djibouti, South Sudan, Central African Republic].
Total Route Cost: 43.

\item $k=5$\textbf{:Southern Africa Cluster.}

PCG: [Kenya, South Africa].
LDTs: [None].
We decide to include South Africa in this cluster and set it as one of the PCG, rather than have it as a unit cluster as shown in Fig. \ref{figZa}. 
Optimal Route: [Zimbabwe, Mozambique, Zambia, Malawi, Democratic Republic of Congo, Burundi, Rwanda, Uganda, Kenya, Tanzania, Comoros, South Africa, Eswatini, Botswana, Angola, Namibia].
Total Route Cost: 37.

\item $k=5$\textbf{:Eastern Africa Cluster.}
PCG: [Mauritius].
LDT: [None].
In the absence of a PCG for this cluster, we chose Mauritius as it has a submarine landing and the highest number of data centres of all the other countries in the cluster.
Optimal Route: [Lesotho, Madagascar, Mauritius, Seychelles, Somalia, Ethiopia].
Total Route Cost: 30.
\end{itemize}

\begin{itemize}
\item $k=6$\textbf{:Western Africa Cluster.}
PCG: [Ghana]. As before no country within the cluster met the requirement for PCG, hence we chose Ghana.
LDTs: [Mali, Western Sahara and Mauritania]. 
Optimal Route: [Burkina Faso, Benin, Togo, Ghana, Cote d'Ivoire, Liberia, Sierra Leone, Guinea, Guinea Bissau, Gambia, Senegal, Cabo Verde, Western Sahara, Mauritania, Mali, Niger].
Total Route Cost= 27.33.

\item $k=6$\textbf{:Northern Africa Cluster.}
PCG: [Algeria].
LDT: [Tunisia].
Egypt and Libya were not included for similar reasons as $k=5$. 
Optimal Route: [Egypt, Libya, Algeria, Morocco, Tunisia].
Total Route Cost = 22.9.

\item $k=6$\textbf{:Central Africa Cluster.}
PCGs: [Cameroon, Djibouti, Kenya, Nigeria].
LDTs: [Chad, Sudan].
Optimal Route: [Equatorial Guinea, Gabonese Republic, Republic of the Congo, Democratic Republic of Congo, Rwanda, Uganda, Kenya, South Sudan, Djibouti, Eritrea, Sudan, Chad, Central African Republic, Cameroon, Nigeria, Sao Tome and Principe].
Total Route Cost: 36.

\item $k=6$\textbf{:Southern Africa Cluster.}
PCG: [South Africa]. As done earlier, South Africa in included in this cluster, to avoid having a unit cluster.
LDT: [None].
Optimal Route: [Malawi, Mozambique, Zimbabwe, Eswatini, South Africa, Botswana, Namibia, Angola, Zambia, Burundi, Tanzania, Comoros].
Total Route Cost: 30.72.

\item $k=6$\textbf{:Eastern Africa Cluster.}
PCG: [Mauritius].
LDT: [None].
Optimal Route: [Seychelles, Somalia, Ethiopia, Lesotho, Madagascar, Mauritius].
Total Route Cost: 36.
\end{itemize}



\begin{itemize}
\item \textbf{AU Western Africa Cluster.}
PCG: [Nigeria].
LDTs: [Mali, Niger].
Optimal Route: [Senegal, Gambia, Guinea Bissau, Guinea, Sierra Leone, Liberia, Cote d'Ivoire, Ghana, Togo, Benin, Nigeria, Niger, Burkina Faso, Mali, Cabo Verde].
Total Route Cost: 26.7.

\item \textbf{AU Northern Africa Cluster.}
PCGs: [Algeria].
LDTs: [Mauritania, Sudan, Tunisia, Western Sahara].
Optimal Route: [Sudan, Egypt, Libya, Tunisia, Algeria, Morocco, Western Sahara, Mauritania].
Total Route Cost: 32.45.

\item \textbf{AU Central Africa Cluster.}
PCGs: [Cameroon].
LDTs: [Chad].
Optimal Route: [Central African Republic, Chad, Cameroon, Equatorial Guinea, Sao Tome and Principe, Gabonese Republic, Republic of the Congo, Democratic Republic of Congo, Burundi].
Total Route Cost: 21.

\item \textbf{AU Southern Africa Cluster.}
PCGs: [South Africa].
LDTs: [None].
Optimal Route: [Angola, Zambia, Malawi, Mozambique, Zimbabwe, Eswatini, Lesotho, South Africa, Botswana, Namibia].
Total Route Cost: 21.5.

\item \textbf{AU Eastern Africa Cluster.}
PCGs: [Kenya, Djibouti].
LDTs: [None].
Optimal Route: [South Sudan, Kenya, Uganda, Rwanda, Tanzania, Comoros, Madagascar, Mauritius, Seychelles, Somalia, Ethiopia, Djibouti, Eritrea].
Total Route Cost: 37.
\end{itemize}

\begin{figure}
\centerline{\includegraphics[scale=0.5]{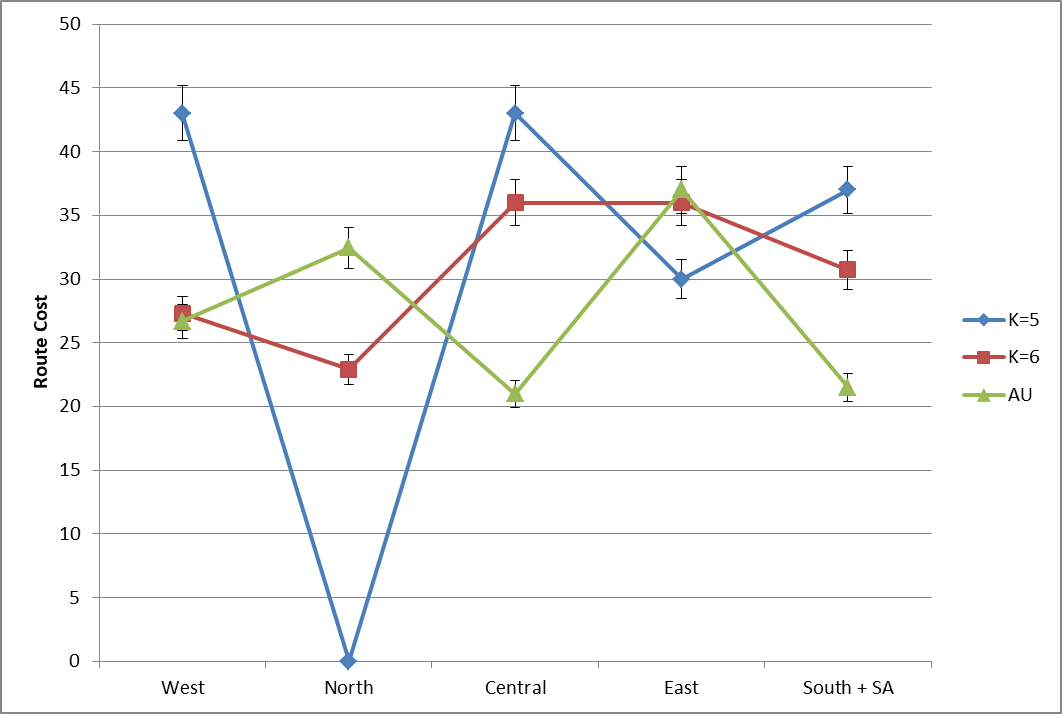}}
\caption{Intra-Cluster Route Cost.}\label{figM2}
\end{figure}

Fig. \ref{figM2} shows the total cost of traversing all countries within each of the clusters for the Multi-feature clustering (both $k=5$ and $k=6$) and the AU’s clustering layout. The Northern cluster for $k=5$ is zero because $k=5$ did not create a Northern cluster – see Fig. \ref{figZa}. Comparatively, Fig. \ref{figM2}, shows that of all the three layouts, the AU's clustering layout resulted in the least traversal cost per cluster. Also routing across the $k=5$ clusters generally cost more compared to $k=6$, except in the East cluster where at 30, it was lower than $k=6$'s 36 and AU's 37.

\subsection{Inter-Cluster Routing}
Having established a path between each cluster and its PCGs, there is the need to create a path across multiple clusters. This is the second (upper) level of our hierarchical clustering model. We again applied the modified ACO-SP algorithm to obtain the optimal route across the various clusters. Though this level represents routing across PCGs and all routes should ideally be desirable, we still included LDTs (which are paths that traverse the Sahara desert). Due to the small number of nodes, we reduced the number iteration to 50 and number of ants to 100. Obtained results are as follows:

\begin{itemize}

\item \textbf{Multi Feature Clustering with $k=5$}
PCGs: [Cameroon, Djibouti, Ghana, Kenya, Nigeria, Mauritius, South Africa]
LDTs: [Egypt, Libya] 
Optimal Route: [Mauritius, South Africa, Cameroon, Nigeria, Ghana, Libya, Egypt, Djibouti, Kenya]
Total Route Cost: 60.7

\item \textbf{Multi Feature Clustering with $k=6$}

PCGs: [Cameroon, Djibouti, Ghana, Kenya, Nigeria, Mauritius, South Africa]
LDTs: [Algeria, Egypt, Libya]
Optimal Route: [Kenya, Djibouti, Egypt, Libya, Algeria, Ghana, Nigeria, Cameroon, South Africa, Mauritius]
Total Route Cost: 63.7

\item \textbf{AU Clustering}
PCGs: [Cameroon, Djibouti, Kenya, Nigeria, South Africa,]
LDTs: [Algeria, Egypt, Libya]
Optimal Route: [Kenya, Djibouti, Egypt, Libya, Algeria, Nigeria, Cameroon, South Africa]
Total Route Cost: 52.6

\end{itemize}

\begin{figure}
\centerline{\includegraphics[scale=0.5]{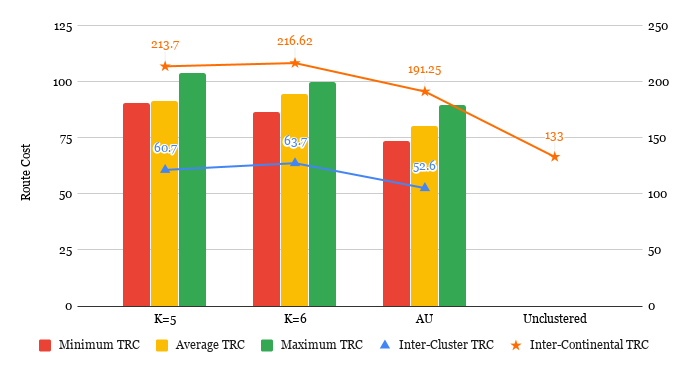}}
\caption{Inter-Cluster Traversal   .\label{figM3}}
\end{figure}

Fig. \ref{figM3} shows the Total Routing Cost (TRC) for traversals across clusters. Of the three models, the AU’s clustering layout resulted in the lowest inter-cluster TRC at 52.6. This was followed by the multi feature clustering with $k$ set to 5, while the option with $k$ set to 6 cost the most to traverse the continent. Similar trends were observed for the average and maximum TRC. The minimum TRC for $k=6$ was however lower than that of clustering with $k=5$. This can be attributed to the fact that $k=6$, created an additional cluster thereby, reducing the number of countries per cluster and by extension the traversal cost of some of its clusters. 

\subsection{Continental Routing Without Clustering}
In this simulation we set each country as a unit entity rather a member of a cluster. Cameroon, Djibouti, Kenya, Nigeria and South Africa were set as PCGs; while least desired trail destinations included: Chad, Mali, Mauritania, Niger, Sudan, Tunisia and Western Sahara, as these are countries within the Sahara desert. Though Algeria, Egypt and Libya are also located in the Sahara, they are not included in the list of undesired nor are they included in the desired list. This is because the influence of the stench multiplier (for undesired paths) and boost multiplier (for desired paths) cancel each other out, hence no point including them in any of the lists. The simulation was run using 100 ants and obtained results are shown with the line graph in Fig. \ref{figM3}.

The line graph on Fig. \ref{figM3} shows that the optimal route to traverse the entire continent would cost a TRC of 133. This was obtained by traversing the continent as follows: Nigeria, Cameroon, Equatorial Guinea, Gabonese Republic, Republic of the Congo, Sao Tome and Principe, Togo, Benin, Ghana, Cote d'Ivoire, Liberia, Sierra Leone, Guinea, Guinea Bissau, Gambia, Senegal, Cabo Verde, Mauritania, Western Sahara, Morocco, Algeria, Tunisia, Libya, Egypt, Sudan, Eritrea, Djibouti, Ethiopia, Somalia, Kenya, Uganda, Rwanda, Burundi, Democratic Republic of Congo, Tanzania, Malawi, Mozambique, Zimbabwe, Zambia, Angola, Namibia, Botswana, South Africa, Lesotho, Eswatini, Madagascar, Mauritius, Comoros, Seychelles, South Sudan, Central African Republic, Chad, Niger, Mali, Burkina Faso.

\begin{figure}
\centerline{\includegraphics[scale=1]{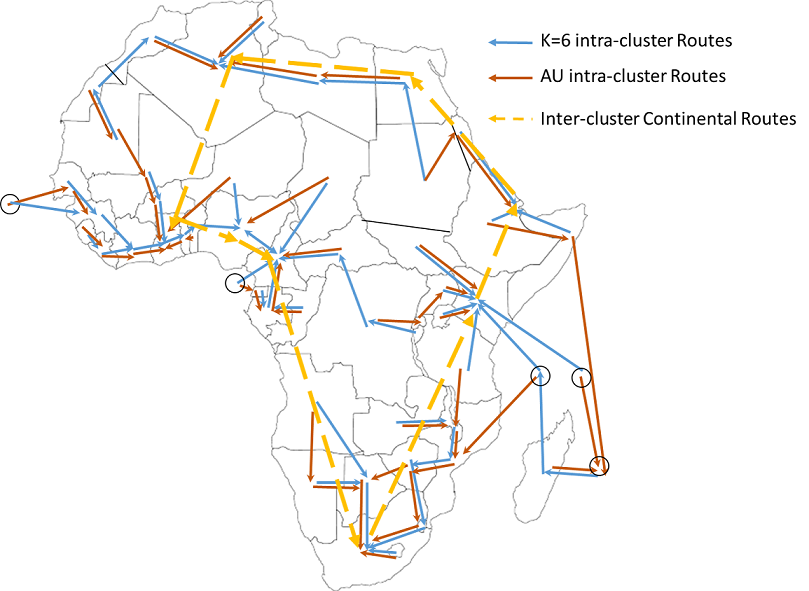}}
\caption{The African Continental Network Layout.\label{figAf3}}
\end{figure}

When compared with the continental traversal obtained using the clustering based approaches, this unclustered approach resulted in the lowest TRC at 133. It was followed by the AU-based hierarchical clustering layout at 191.25, while the combined feature clustering with $k$ set to 6 was the least effective approach requiring the highest TRC at 216.62. A summary of the routing costs is shown on Table \ref{tab5} and depicted in Fig. \ref{figAf3}.

\begin{table}

\caption{Summary of Routing Costs}
\label{tab5}
\begin{scriptsize}
\begin{tabular}{p{1.15cm}|p{0.8cm}|p{0.5cm}|p{0.8cm}|p{0.5cm}|p{0.75cm}|p{0.5cm}|p{0.8cm}}
\hline
& \multicolumn{2}{|p{1.5cm}|}{Multi Feature Clustering $(K=5)$} & 
\multicolumn{2}{|p{1.5cm}|}{Multi Feature Clustering $(K=6)$} &
\multicolumn{2}{|p{1.5cm}|}{AU Clustering} & Unit Clustering
\\ \hline

&	Country Count &	Route Cost & Country Count &	Route Cost & Country Count & Route Cost	
\\ \hline

Western Cluster & 18 &	43 & 16 & 27 & 15 &	27 & NA \\ \hline
Northern Cluster &	0 &	0 &	5 &	23 & 8 & 32	 & NA \\ \hline
Central Cluster & 15 & 43 &	16 & 36 & 9 & 21 & NA \\ \hline
Eastern Cluster & 6 & 30 & 6 & 36 &	13 & 37 & NA
\\ \hline
Southern Cluster & 16 & 37 & 12 & 31 & 10 & 21	& NA \\ \hline
Intra-Cluster \par Total & 	55 & 153 & 55 &	152 & 55 & 138 &	55 \\ \hline
Inter-Cluster \par Total & 9 (PCGs) & 61 & 10  (PCGs) & 64 & 8 (PCGs) &	54 &	8 (PCGs) \\ \hline
Continental Total \par (Intra + Inter) & & 214 & & 216 & & 191 & 131 
\\ \hline

\end{tabular}
\end{scriptsize}
\label{tab5}
\end{table}

\section{Implementation Considerations} \label{s6}
In this section, we briefly discuss a potential use case of the model developed in this work; as well as some challenges that might hinder its implementation.

\subsection{Potential Use Case}
The relevance or potential use case of this work, is in meeting the African Sustainable Development Goals (SDG) as defined in \cite{sdg}. Of particular concern are  those of: i. good health and well being (through collaborative healthcare \cite{FedCloud, Mandava}); ii. Industry, innovation and infrastructure (through scientific research and high performance computing such as Cloud computing); and iii.) Sustainable cities and communities (smart and eco-cities \cite{sm}). These three goals are hinged on technologies of the fourth industrial revolution. An underlying network capable of supporting high bandwidth and data throughput is paramount to achieving this. Our Africa 3 model proposes a means of efficiently architecting such a network.

\subsection{Open Challenges}
Thus far, deploying a pan African network for collaboration has remained elusive. A number of challenges are associated with implementing such a network, some of which are as follows:

\begin{itemize}
\item \textbf{Policies:}
The policies (or lack of) are a major deterrent to collective growth of Africa. A number of policies on Agriculture and food security, health and education are in place but have remained dormant or ill-implemented. From a data and ICT standpoint, the Protection of Personal Information Act and data sovereignty \cite{popi, sello, up} are fundamental issues that need to be addressed. They are concerned with data legislation on where, who and how data should be stored or accessed across national boundaries. These need to be clearly agreed upon before a truly collaborative data network, such as Africa 3 can be implemented.

\item \textbf{Technical Expertise:}
Beyond policies, technical experts would be needed to design, deploy and operate a continental network such as Africa 3. Unfortunately, as of today, there is still a wide gap in technical know-how between Africa and the developed world. This is especially true for ICT and related fields. In order to achieve a sustainable continental network, Africans need to be up-skilled in emerging fields such areas next generation networks, Cloud computing and data engineering. There is also the need to focus on Cyber security and threats, as attacks on a continental network could have far reaching consequences.

\item \textbf{Electricity:} Without a doubt, electricity is a fundamental requirement for any data network. Though at a continental level, one would assume that electricity would be available to power Africa 3; the actual problem lies at the edges or peripherals of this network. Here, the term edges refer to the individual countries connected to Africa 3, and which serve as "data prosumers". As of today, electricity remains one of the biggest bottlenecks to sustainable development in many edges. Hence, for a continental network such as Africa 3 to be truly useful, African countries would require widespread energy reforms. This includes exploring alternate, renewable and sustainable sources of electricity such as wind, solar, wave and hydro.

\item \textbf{Poverty:}
Though Africa 3 is being proposed to help accelerate development across Africa, the model in itself has to be viable / self-sustainable. Users therefore have to pay to use it. Sadly, over 60\% of the sub-Saharan populace still live in poverty \cite{who14}. Many families are unable to afford good food, healthcare or education for their children. Poverty translates to little purchasing power, weak economies and ultimately little or no funds to sustain the continental network. If deployed as a non-commercial entity and left unfunded, such continental network would collapse / die-off at its infancy.

\end{itemize}

\section{Conclusion} \label{s7}
Reports of the United Nations on global human development index, shows that more than 75\% of countries in Africa are underdeveloped. These countries lack vital infrastructure for health care, education, transportation, electricity etc. This situation is further exacerbated by the high level of poverty in many of these countries. Collaboration enabled by Information and communication Technology (ICT) could prove to be a vital tool in fast-tracking development across Africa. In achieving this, an underlying robust intra-continental network structure must first be put in place; upon which concepts such as tele-medicine, remote learning, e-governance, smart cities and federated Cloud processing and storage can be built. The intricacies of such a network are often enormous as factors such as geo-location, population size, socio-economic and politics need to be considered. These therefore complicate the process. In this work, we proposed a network model for inter-connecting African countries. This two-layered hierarchical model takes into consideration the population size, the geographical locations, the number of independent sub-marine cable landings and the number of high performance computing infrastructure available in each African country into consideration. At the first layer, African countries are grouped into functional clusters. Four clustering algorithms were considered, viz. K-Means, K-Medoids, Hierarchical clustering and OPTICS-Xi. These were bench-marked against the Africa Union’s clustering (AU) and obtained results show that the K-Medoids gave results closest to the AU. Upon successfully clustering the countries, a modified variant of the Ant Colony Optimization algorithm called Ant Colony Optimization with Stench Pheromone was used to obtain least cost network paths within and between the clustered. Stench Pheromones were used to dissuade paths that traversed the Sahara desert. Simulation results show that indeed an optimal network path can be obtained that traversed all African nations. Obtained results also show that it is cheaper to design such an African network based on the Africa Union’s clustering layout. Implementation challenges of this network model in Africa were then discussed.

Though a robust network model for Africa was proposed in this work; the authors have assumed that perfect line of sight exists when linking countries. Furthermore, ground elevation, weather conditions and geo-political constraints such as language diversity and trans-border rifts were not considered. Finally, though a network has been proposed, the exact technology in terms of transmission media was also not considered. These are potential areas that might be considered for future works.

\section{Acknowledgement}
The authors acknowledge Prof. Biswanath Mukerjee and Dr. Clement Nyirenda for their review of the paper and comments, both of which helped improved the work.

\section{Appendix} \label{apx}

\begin{table}
\centering
\caption{Routing Path Per Country Using K=5 }
\label{tab6}
\begin{scriptsize}
\begin{tabular}{p{1.5cm}|p{1.5cm}|p{8.5cm}|l}
\hline
\textbf{Source} & \textbf{Destination (PCGs)} & \textbf{Path} & \textbf{Cost}
\\ \hline
Benin	& Algeria &	Benin, Togo, Ghana, Burkina Faso, Mali, Mauritania, Western Sahara, Morocco, Algeria &	20.77 \\ \hline
Burkina Faso &	Algeria	& Burkina Faso, Mali, Mauritania, Western Sahara, Morocco, Algeria &	17.10 \\ \hline
Cabo Verde & Algeria &	Cabo Verde, Gambia,Guinea Bissau,Senegal, Mauritania,Western Sahara,Morocco,Algeria &	22.15 \\ \hline
Cote d’Ivoire &	Algeria &	Cote d'Ivoire,Liberia,Sierra Leone,Guinea,Guinea Bissau,Gambia,Senegal,Mauritania,Western Sahara,Morocco,Algeria &	22.44 \\ \hline
Gambia	& Algeria &	Gambia, Senegal	 Guinea Bissau, Guinea, Sierra Leone,	Liberia, Cote d'Ivoire, Ghana, Togo, Benin, Burkina Faso, Mali, Mauritania, Western Sahara, Morocco	 Algeria &	28.07 \\ \hline
Ghana	& Algeria &	Ghana, Togo, Benin, Burkina Faso, Mali, Mauritania, Western Sahara, Morocco, Algeria &	21.38 \\ \hline
Guinea &	Algeria &	Guinea, Sierra Leone,	 Liberia,	 Cote d'Ivoire,	 Ghana,	 Togo,	 Benin,	 Burkina Faso,	 Mali,	 Mauritania,	 Western Sahara, Morocco,	 Algeria &	25.62 \\ \hline
Guinea Bissau &	Algeria &	Guinea Bissau,	 Gambia, Senegal, Mauritania,	Western Sahara, Morocco, Algeria &	18.12 \\ \hline
Liberia	& Algeria &	Liberia,	 Sierra Leone,	 Guinea,	 Guinea Bissau,	 Gambia,	 Senegal,	 Mauritania,	 Western Sahara,	 Morocco, Algeria &	22.03 \\ \hline
Mali &	Algeria &	Mali,	 Mauritania,	 Western Sahara,	 Morocco, Algeria &	13.61 \\ \hline
Mauritania & Algeria &	Mauritania,	 Western Sahara, Morocco, Algeria &	11.77 \\ \hline
Morocco	& Algeria &	Morocco, Algeria &	6.53 \\ \hline
Senegal	& Algeria &	Senegal,	 Gambia,	 Guinea Bissau,	 Guinea,	 Sierra Leone,	 Liberia, Cote d'Ivoire, Ghana,	 Togo, Benin,	 Burkina Faso, Mali, Mauritania,	 Western Sahara,	 Morocco,	 Algeria &	27.32 \\ \hline
Sierra Leone &	Algeria &	Sierra Leone,	 Guinea, Guinea Bissau,	 Gambia,	 Senegal, Mauritania,	 Western Sahara,	 Morocco,	 Algeria &	20.64 \\ \hline
Togo &	Algeria &	Togo,	 Ghana,	 Benin,	 Burkina Faso,	 Mali,	 Mauritania,	 Western Sahara,	 Morocco, Algeria & 21.72 \\ \hline
Tunisia &	Algeria &	Tunisia, Algeria & 5.90 \\ \hline
Western Sahara & Algeria &	Western Sahara,	 Mauritania, Senegal,	 Gambia,	 Guinea Bissau,	 Guinea,	 Sierra Leone,	 Liberia,	 Cote d'Ivoire,	 Ghana,	 Togo,	 Benin,	 Burkina Faso,	 Mali,	 Algeria &	23.74 \\ \hline

C.A.R & Cameroon &	Central African Republic, Cameroon &	4.41 \\ \hline
Chad	& Nigeria or Cameroon &	Chad, Nigeria or Cameroon &	7.71 \\ \hline
Egypt	& Djibouti &	Egypt,	Eritrea, Djibouti &	13.09 \\ \hline
Equatorial Guinea &	Cameroon &	Equatorial Guinea, Gabonese Republic, Republic of the Congo	 Cameroon &	5.94\\ \hline
Eritrea	& Djibouti &	Eritrea, Djibouti	& 3.33\\ \hline
Gabonese Republic &	Cameroon &	Gabonese Republic, Equatorial Guinea, Cameroon &	4.79\\ \hline
Libya &	Nigeria	 & Libya,	 Niger,	 Nigeria &	14.39\\ \hline
Niger &	Nigeria & Niger,	 Nigeria &	6.17\\ \hline
Rep. of Congo &	Cameroon &	Republic of the Congo,	Gabonese Republic,	 Equatorial Guinea, Cameroon &	5.95\\ \hline
Sao Tome and Principe & Cameroon &	Sao Tome and Principe, Gabonese Republic,	 Equatorial Guinea,	Cameroon &	6.52\\ \hline
South Sudan	& Djibouti & South Sudan, Djibouti &	7.61\\ \hline
Sudan & Cameroon &	Sudan,	 Libya,	 Chad,	 Cameroon &	21.84\\ \hline

Angola	& South Africa &	Angola,	Namibia, Botswana, South Africa	& 14.71 \\ \hline
Botswana	& South Africa &	Botswana,	 South Africa &	5.37 \\ \hline
Burundi	& Kenya	& Burundi, Rwanda, Uganda, Kenya & 6.28 \\ \hline
Comoros	& Kenya	& Comoros,	 Tanzania,	 Burundi,	 Rwanda,	 Uganda,	 Kenya &	13.44 \\ \hline
DRC	& Kenya &	Democratic Republic of Congo,	 Burundi, Rwanda, Uganda, Kenya &	9.71 \\ \hline
Eswatini &	South Africa &	Eswatini,  South Africa &	4.42\\ \hline
Malawi &	South Africa &	Malawi,	 Mozambique,	 Zimbabwe,	 Botswana,	 South Africa &	14.69\\ \hline
Mozambique &	South Africa &	Mozambique,	 Malawi,	 Zambia,	 Zimbabwe,	 Eswatini, 	 South Africa &	15.81\\ \hline
Namibia	& South Africa &	Namibia,	 Botswana,	 South Africa &	7.05\\ \hline
Rwanda &	Kenya &	Rwanda	, Burundi, Kenya &	5.72\\ \hline
Tanzania &	Kenya & Tanzania, Rwanda, Burundi, Kenya &	8.08\\ \hline
Uganda &	Kenya &	Uganda, Kenya &	3.30\\ \hline
Zambia &	Kenya	& Zambia, DRC, Burundi, Rwanda, Burundi, Uganda, Kenya &	15.31\\ \hline
Zimbabwe &	South Africa &	Zimbabwe, Eswatini, South Africa &	9.03\\ \hline
Ethiopia &	Mauritius	& Ethiopia, Somalia, Mauritius &	23.63\\ \hline
Lesotho	& Mauritius &	Lesotho	, Mauritius &	17.88\\ \hline
Madagascar	& Mauritius &	Madagascar, Mauritius &	6.31\\ \hline
Seychelles	& Mauritius &	Seychelles, Mauritius &	11.39\\ \hline
Somalia &	Mauritius &	Somalia,	 Ethiopia, Mauritius & 23.63\\ \hline

\end{tabular}
\end{scriptsize}
\end{table}

\begin{table*}[]
\centering
\caption{Routing Path Per Country Using K=6}
\label{tab7}
\begin{scriptsize}
\begin{tabular}{l|l|p{6cm}|l}
\hline
\textbf{Source} & \textbf{Destination (PCGs)} & \textbf{Path} & \textbf{Cost}
\\ \hline
Benin	& Ghana &	Benin, Togo, Ghana &	2.46 \\ \hline
Burkina Faso	& Ghana &	Burkina Faso, Ghana &	3.21 \\ \hline
Cabo Verde	& Ghana &	Cabo Verde,	Senegal, Gambia, Guinea Bissau, Guinea, Liberia, Cote d'Ivoire,	 Ghana & 	15.94 \\ \hline
Cote d’Ivoire	& Ghana &	Cote d'Ivoire, Ghana &	2.56 \\ \hline
Gambia	& Ghana &	Gambia,	 Senegal,	 Guinea Bissau,	 Guinea, Sierra Leone,	 Liberia, Cote d'Ivoire,	 Ghana &	11.22 \\ \hline
Guinea & Ghana & Guinea, Sierra Leone, Liberia, Cote d'Ivoire, Ghana &	7.10 \\ \hline
Guinea Bissau & Ghana &	Guinea Bissau, Gambia, Senegal, Sierra Leone, Liberia, Cote d'Ivoire, Ghana & 10.79 \\ \hline
Liberia	& Ghana &	Liberia, Cote d'Ivoire, Ghana & 4.81 \\ \hline
Mali & Ghana &	Mali, Burkina Faso, Ghana & 6.99 \\ \hline
Mauritania & Ghana & Mauritania,	Mali, Burkina Faso, Ghana & 10.84 \\ \hline
Niger &	Ghana &	Niger, Ghana &	9.12 \\ \hline
Senegal	& Ghana &	Senegal, Gambia, Guinea Bissau, Guinea, Sierra Leone, Cote d'Ivoire, Ghana &	10.44 \\ \hline
Sierra Leone & Ghana &	Sierra Leone, Guinea, Liberia, Cote d'Ivoire, Ghana &	7.68\\ \hline
Togo & Ghana &	Togo, Ghana &	1.37\\ \hline
Western Sahara & Ghana & Western Sahara, Mauritania,Mali, Burkina Faso, Ghana &	13.86\\ \hline

Egypt	& Algeria &	Egypt, Libya, Algeria &	15.72 \\ \hline
Libya	& Algeria &	Libya, Algeria &	8.75 \\ \hline
Morocco	& Algeria &	Morocco, Algeria &	6.53 \\ \hline
Tunisia & Algeria &	Tunisia, Algeria &	5.90 \\ \hline
Chad &	Nigeria &	Chad, Nigeria  & 7.71 \\ \hline
C.A.R &	Cameroon &	Central African Republic, Cameroon &	4.41\\ \hline
DRC & Kenya &	DRC, Rwanda, Uganda, Kenya &	9.23\\ \hline
Equatorial Guinea &	Cameroon & Equatorial Guinea, Gabonese Republic, Republic of the Congo,	 Cameroon &	5.94\\ \hline
Eritrea &	Djibouti & Eritrea, Djibouti &	3.33\\ \hline
Gabonese Republic	& Cameroon & Gabonese Republic, Equatorial Guinea, Cameroon &	4.79\\ \hline
Rep. of Congo &	Cameroon & Republic of the Congo, Gabonese Republic, Equatorial Guinea,	 Cameroon &	5.95\\ \hline
Rwanda &	Kenya & Rwanda, Kenya &	4.97 \\ \hline
Sao Tome and Principe & Cameroon & Sao Tome and Principe, Gabonese Republic, Equatorial Guinea, Cameroon &	6.52\\ \hline
South Sudan & Kenya & South Sudan, Kenya &	6.52\\ \hline
Sudan &	Djibouti & Sudan, Eritrea, Djibouti &	8.09\\ \hline
Uganda & Kenya & Uganda, Kenya &	3.30\\ \hline

Angola &	South Africa &	Angola,	Namibia, Botswana, South Africa	& 14.71 \\ \hline
Botswana & South Africa &	Botswana,	 South Africa &	5.37 \\ \hline
Burundi & South Africa &	Burundi,	 Tanzania,	 Malawi,	 Mozambique,	 Zimbabwe, Botswana,	 South Africa or Burundi, Tanzania, Malawi,	 Mozambique, Zimbabwe, Eswatini, South Africa &	22.0 \\ \hline
Comoros &	South Africa & Comoros,	 Mozambique,	 Malawi,	 Zambia,	 Eswatini, 	 South Africa or Comoros,	 Malawi,	 Mozambique,	 Zimbabwe,	 Botswana,	 Namibia,	 South Africa &	21.68 \\ \hline
Eswatini &	South Africa &	Eswatini, South Africa &	4.41\\ \hline
Malawi &	South Africa &	Malawi,	 Mozambique, Zimbabwe, Botswana,South Africa &	14.47\\ \hline
Mozambique & South Africa & Mozambique, Zimbabwe, Botswana, South Africa & 11.86 \\ \hline
Namibia & South Africa & Namibia, Botswana, South Africa & 8.05 \\ \hline
Tanzania & South Africa & Tanzania,	 Burundi, Zambia, Zimbabwe, Botswana, South Africa &	21.17\\ \hline
Zambia	& South Africa & Zambia, Malawi, Mozambique, Zimbabwe, Eswatini, South Africa & 15.95\\ \hline
Zimbabwe & South Africa & Zimbabwe, Eswatini, South Africa & 9.03\\ \hline
Ethiopia & Mauritius & Ethiopia, Somalia, Mauritius & 23.63\\ \hline
Lesotho & Mauritius & Lesotho, Mauritius &	17.88\\ \hline
Madagascar & Mauritius & Madagascar, Mauritius & 6.31\\ \hline
Seychelles & Mauritius & Seychelles, Mauritius & 11.39\\ \hline
Somalia & Mauritius & Somalia, Ethiopia, Mauritius & 23.63\\ \hline

\end{tabular}
\end{scriptsize}
\end{table*}

\begin{table*}[]
\centering
\caption{Routing Path Per Country Using AU's model}
\label{tab8}
\begin{scriptsize}
\begin{tabular}{l|l|p{6cm}|l}

\hline
\textbf{Source} & \textbf{Destination (PCGs)} & \textbf{Path} & \textbf{Cost}
\\ \hline
Benin	& Nigeria &	Benin, Nigeria &	2.36  \\ \hline
Burkina Faso & Nigeria & Burkina Faso, Ghana, Togo, Benin, Nigeria &	7.90 \\ \hline
Cabo Verde & Nigeria &	Cabo Verde,	 Guinea, Sierra Leone, Liberia, Cote d'Ivoire,	Ghana, Togo, Benin, Nigeria & 20.96 \\ \hline
Cote d’Ivoire & Nigeria & Cote d'Ivoire, Ghana,	 Togo,	Benin, Burkina Faso,	 Nigeria &	11.36 \\ \hline
Gambia	& Nigeria &	Gambia, Senegal, Guinea Bissau, Guinea, Sierra Leone, Liberia, Cote d'Ivoire, Ghana, Togo, Benin, Nigeria or Gambia, Senegal, Guinea Bissau, Sierra Leone, Guinea, Cote d'Ivoire, Ghana, Togo, Benin, Nigeria &	16.77\\ \hline
Ghana &	Nigeria & Ghana, Togo, Benin, Nigeria &	6.07\\ \hline
Guinea & Nigeria &	Guinea, Sierra Leone, Liberia, Cote d'Ivoire, Ghana, Togo, Benin, Nigeria &	12.89\\ \hline
Guinea Bissau	& Nigeria &	Guinea Bissau, Gambia, Senegal, Guinea, Sierra Leone, Liberia, Cote d'Ivoire, Ghana, Togo, Benin, Nigeria &	16.76\\ \hline
Liberia & Nigeria &	Liberia, Cote d'Ivoire, Ghana, Togo, Benin, Nigeria &	10.80\\ \hline
Mali & Nigeria & Mali, Burkina Faso, Ghana, Togo, Benin, Nigeria &	11.18\\ \hline
Niger &	Nigeria & Niger, Nigeria &	6.17\\ \hline
Senegal	& Nigeria &	Senegal, Gambia, Guinea Bissau,	Guinea, Sierra Leone, Liberia, Ghana, Togo, Benin, Nigeria &	16.33\\ \hline
Sierra Leone & Nigeria & Sierra Leone, Guinea, Liberia, Cote d'Ivoire, Ghana, Togo, Benin, Nigeria &	13.61\\ \hline
Togo & Nigeria & Togo, Benin, Nigeria & 4.71\\ \hline

Burundi	& Cameroon & Burundi, Democratic Republic of Congo, Cameroon &	12.08\\ \hline
C.A.R & Cameroon &	Central African Republic, Cameroon &	4.41\\ \hline
Chad & Cameroon & Chad, Cameroon &	7.71\\ \hline
DRC & Cameroon & Democratic Republic of Congo, Burundi, Central African Republic,	Cameroon &	12.91\\ \hline
Equatorial Guinea &	Cameroon &	Equatorial Guinea, Gabonese Republic, Republic of the Congo,	 Cameroon & 5.94\\ \hline
Gabonese Republic &	Cameroon &	Gabonese Republic, Equatorial Guinea, Sao Tome and Principe, Cameroon &	6.85\\ \hline
Rep. of Congo	& Cameroon & Republic of the Congo, Gabonese Republic, Equatorial Guinea,	 Cameroon &	5.95\\ \hline
Rwanda & Kenya & Rwanda, Kenya & 4.97\\ \hline
Sao Tome and Principe & Cameroon &	Sao Tome and Principe,	Gabonese Republic, Equatorial Guinea, Cameroon & 6.52\\ \hline
Angola & South Africa &	Angola, Botswana, Namibia, South Africa &	15.97\\ \hline
Botswana & South Africa & Botswana, South Africa & 5.37\\ \hline
Eswatini & South Africa & Eswatini, South Africa &	4.41\\ \hline
Lesotho & South Africa & Lesotho, South Africa & 2.24\\ \hline
Malawi & South Africa &	Malawi,	Mozambique,	 Zimbabwe	 Botswana,	 South Africa &	14.47\\ \hline
Mozambique	& South Africa &	Mozambique, Zimbabwe, Botswana, South Africa &	11.86\\ \hline
Namibia &	South Africa & Namibia, Botswana, South Africa &	8.05\\ \hline
Zambia & South Africa &	Zambia, Malawi,	Mozambique, Zimbabwe, Eswatini, South Africa & 15.95\\ \hline
Zimbabwe & South Africa &	Zimbabwe, Eswatini, South Africa &	9.03\\ \hline
Comoros & Kenya & Comoros, Kenya &	9.70\\ \hline
Ethiopia & Djibouti & Ethiopia, Djibouti &	2.97\\ \hline
Eritrea & Djibouti & Eritrea, Djibouti & 3.33\\ \hline
Madagascar & Kenya & Madagascar, Comoros, Tanzania, Kenya & 16.51\\ \hline
Mauritius &	Kenya &	Mauritius, Madagascar, Comoros, Tanzania, Kenya &	21.44\\ \hline
Rwanda & Kenya & Rwanda, Uganda, Kenya & 5.53\\ \hline
Seychelles & Kenya & Seychelles, Kenya & 10.76\\ \hline
Somalia & Djibouti or Kenya & Somalia, Djibouti or Somalia, Kenya & 5.4 or 5.5\\ \hline
South Sudan & Kenya &	South Sudan, Kenya &	6.58\\ \hline
Tanzania &	Kenya &	Tanzania, Kenya &	5.47\\ \hline
Uganda & Kenya & Uganda, Kenya &	3.30 \\ \hline

\end{tabular}
\end{scriptsize}
\end{table*}


\begin{thebibliography}{00}
\bibitem{FedCloud} Ajayi, O., Bagula, A.,  Ma, K. (2019). "Fourth Industrial Revolution for Development: The Relevance of Cloud Federation in Healthcare Support," in IEEE Access, 7, pp. 185322-185337.

\bibitem{dcmap} DataCentreMap. Retrieved from: \url{datacentermap.com}. Accessed 21/03/2020

\bibitem{africa2} Huang, A., Sarkar, S., Mukherjee, B. (2008). Africa Two: A Proposal for a Concentric Two-Ring Network for the African Continent. in IEEE J. on Selected Areas in Communications, 26(6), pp. 2-11.

\bibitem{afPop} Worldometers (2019). African Countries by Population. \url{worldometers.info/population/countries-in-africa-by-population/}. Accessed 9/11/2019.

\bibitem{GADM} GADM Maps and Data. Retrieved from: \url{gadm.org/index.html}. Accessed 19/03/2020.

\bibitem{diminico} DiMinico, C., Jew, J. (2016). Telecommunications infrastructure standard for data centers ANSI/TIA-942. 

\bibitem{AU} African Union (2019). Member States. Retrieved from: \url{au.int/en/member-states/countryprofiles2}. Accessed 10/11/2019

\bibitem{SWAC} West Africa Brief, The Six Regions of the African Union. Retrieved from: \url{ http://www.west-africa-brief.org/content/en/six-regions-african-union}. Accessed 26/02/2020.

\bibitem{Montreal} Apparicio, P., Riva, M., Seguin, A. (2015) A comparison of two methods for classifying trajectories: a case study on neighborhood poverty at the intra-metropolitan level in Montreal ", Cybergeo: European J. of Geography, Espace, Society, Territory, document 727. doi:10.4000/cybergeo.27035

\bibitem{dbscan} Ester, M., Kriegel, H.; Sander, J., Xu, X. (1996). A density-based algorithm for discovering clusters in large spatial databases with noise. Proc.2nd Intl. Conf. on Knowledge Discovery and Data Mining (KDD-96). AAAI Press. pp. 226–231.

\bibitem{hac}  Murtagh, F., Contreras, P. (2011). Methods of hierarchical clustering. arXiv preprint arXiv:1105.0121.

\bibitem{Seacom} Seacom Network. Retrieved from: \url{www.seacom.com/network}. Accessed 11/03/20
 
\bibitem{c1} Xu D, Tian Y (2015). A comprehensive survey of clustering algorithms. Annals of Data Science, 2(2):165-93.

\bibitem{c2}  Cheng, D., Zhu, Q., Huang, J., Wu, Q., Yang, L. (2019). A hierarchical clustering algorithm based on noise removal. Intl. J. of Machine Learning and Cybernetics, 10(7): 1591-602.

\bibitem{c3} Chen, Y., Tang, S., Pei, S., Wang, C., Du, J., Xiong, N. (2017). DHeat: A density heat-based algorithm for clustering with effective radius. IEEE Trans. on Systems, Man, and Cybernetics: Systems, 48(4):649-60.

\bibitem{c4} Jiang, J., Chen, Y., Meng, X., Wang, L., Li, K. (2019). A novel density peaks clustering algorithm based on k nearest neighbors for improving assignment process. Physica A: Statistical Mechanics and its Applications, 523:702-13.

\bibitem{c5} Chen, Y., Tang, S., Bouguila, N., Wang, C., Du, J., Li, H. (2018). A fast clustering algorithm based on pruning unnecessary distance computations in DBSCAN for high-dimensional data. Pattern Recognition, 83:375-87.
 
\bibitem{SAT3} Africa Network Connections - SAT3 / WASC / SAFE. Retrieved from: \url{cmcnetworks.net/sat3-wasc-safe.html}. Accessed 11/03/20

\bibitem{sub} Submarine Cables. Retrieved from: \url{submarinecablemap.com}. Accessed on 11/03/20

\bibitem{aco} Dorigo M., Birattari M., Stutzle T. (2006). Ant colony optimization. IEEE computational intelligence magazine,1(4),28-39.

\bibitem{kmedoid} Kaur, N., Kaur, U., Singh, D. (2014). K-Medoid clustering algorithm-a review. Intl. J. of Computer Application and Tech. (IJCAT), 1(1), 2349-1841.

\bibitem{km} Morissette, L., Chartier, S. (2013). The k-means clustering technique: General considerations and implementation in Mathematica. Tutorials in Quantitative Methods for Psychology, 9(1), 15-24. 

\bibitem{euclidean} Dokmanic, I., Parhizkar, R., Ranieri, J., Vetterli, M. (2015). Euclidean distance matrices: essential theory, algorithms, and applications. IEEE Signal Processing Magazine, 32(6), 12-30.

\bibitem{haver} Robusto, C. (1957). The cosine-haversine formula. The American Mathematical Monthly, 64(1), 38-40.

\bibitem{optix} Ankerst, M., Breunig, M.,  Kriegel, H., Sander, J. (1999). OPTICS: ordering points to identify the clustering structure. SIGMOD Rec. 28(2):49–60. doi: 10.1145/304181.304187

\bibitem{rdrr} Optics. Retrieved from: \url{rdrr.io/cran/dbscan/man/optics.html}. Accessed 10/03/20 

\bibitem{nom1} Patro, S., Sahu, K. (2015). Normalization: A Preprocessing Stage. Department of Comp. Sci. Engineering and Intelligent Transport (CES \& IT), Veer SurendraSai University of Technology (VSSUT), Burla, Odisha, India.

\bibitem{nom2} Mohamad, I, Usman, D. (2013). Standardization and Its Effects on K-means Clustering Algorithm. Research J. of Applied Sciences, Engineering and Technology. 6(17), pp 3299-3303. DOI:10.19026/rjaset.6.3638
  
\bibitem{elki} Schubert, E., Zimek, A. (2019). ELKI: A large open-source library for data analysis-ELKI Release 0.7. 5" Heidelberg". arXiv preprint arXiv:1902.03616.

\bibitem{scikit} Pedregosa, F., Varoquaux, G., Gramfort, A., Michel, V., Thirion, B., Grisel, O. and Vanderplas, J. (2011). Scikit-learn: Machine learning in Python. J. of Machine Learning research, 12, 2825-2830.

\bibitem{sklearn} Scikit Yellowbrick, Elbow Method. Retrieved from: \url{scikit-yb.org/en/latest/api/cluster/elbow.html}. Accessed 18/03/2020

\bibitem{ahn} Caro, F., Gambardella, L. (2005). Anthocnet:An adaptive nature-inspired algorithm for routing in mobile adhoc networks.Special Issue on Self-Organisation in Mobile Networking, 16:443–45.

\bibitem{antnet}Caro, G., Dorigo, M. (1998). Ant colonies for adaptive routing in packet-switched communications networks. In Proc. of ACM Intl. Conf. on Parallel Problem Solving from Nature, pp. 673–682

\bibitem{mobile} Kamali, S., Opatrny, J. (2008). A position Based Ant Colony Routing Algorithm for Mobile Ad-hoc Networks. J. of Networks, 3(4), 31-41, Academic Publisher 

\bibitem{acr} Cong, Z., De Schutter, B., Babuška, R. (2013). Ant Colony Routing algorithm for freeway networks. Transportation Research Part C: Emerging Technologies, 37, 1–19. doi:10.1016/j.trc.2013.09.008 

\bibitem{aco-sp} Cong, Z., De Schutter, B., and Babuska, R., 2011. A new ant colony routing approach with a trade-off between system and user optimum. In proc. of IEEE Intl. Conf. on Intelligent Transportation Systems. Washington, DC, pp. 1369-1374.

\bibitem{bell} Bell, J. McMullen, P. (2004). Ant colony optimization techniques for the vehicle routing problem. Adv. Engineering Informatics, 18, 41–48. Elsevier doi:10.1016/j.aei.2004.07.001

\bibitem{uno} United Nations Development Program (2019). Human Development Indices and Indicators. Human Development Report 2019. HDRO (Human Development Report Office) United Nations Development Programme. pp. 22–25. 

\bibitem{AfREN} Foley, M. (2016). The Role and Status of National Research and Education Networks in Africa.

\bibitem{Mandava} Mandava, M., Lubamba, C., Ismail, A., Bagula, H., Bagula, A. (2016). Cyber- Healthcare for Public Healthcare in the Developing World, in proc. of IEEE Symp. on Computers \& Communication (ISCC), Messina Italy, 14-19.

\bibitem{Adams12} Adams, J.  (2012). Collaborations: The Rise of Research Networks. Nature, 490, 335-336

\bibitem{africa1} Merra, W., Schesser, J. (1996). Africa ONE: The Africa Optical Network. IEEE Communications Magazine, 50-57.

\bibitem{mell} Mell, P., Grance, T. (2011). The NIST Definition of Cloud Computing. NIST Special
Publication 800-145 Whitepaper (p.7)

\bibitem{djamen} Djamen, J., Ramazani, D., Some, S. (1995) Networking in Africa: An unavoidable evolution towards the Internet. Technical report 937 of the Departement d'Informatique et de Recherche Operationnelle (IRO), Universite de Montreal.

\bibitem{is} Rey-Moreno, C. (2017). Supporting the creation and scalability of affordable access solutions: Understanding Community Networks in Africa. Retrieved from: \url{internetsociety.org/wp-content/...May2017-1.pdf}. Accessed 1/4/2020.

\bibitem{chavula} Chavula, J., Feamster, N.,  Bagula, A., Suleman, H. (2014). Quantifying the effects of circuitous routes on the latency of intra-Africa internet traffic: a study of research and education networks. Intl. Conf. on e-Infrastructure and e-Services for Developing Countries, 64-73.  

\bibitem{popi} De Bruyn, M. (2014). The protection of personal information (POPI) act: impact on South Africa. Intl. Business \& Economics Research J., 13(6):1315

\bibitem{sello} Sello, P., Bagula, A., Ajayi, O. (2019). Laws and Regulations on Big Data Management: The Case of South Africa. In Intl. Conf. on e-Infrastructure and e-Services for Developing Countries (pp. 169-179). Springer, Cham.

\bibitem{up}Kushwaha, N., Roguski, P., Watson, B. W. (2020, May). Up in the Air: Ensuring Government Data Sovereignty in the Cloud. In 2020 12th International Conference on Cyber Conflict (CyCon) (Vol. 1300, pp. 43-61). IEEE.

\bibitem{sdg} Henao L, Moyer L, Namakula P. (2017). Africa 2030: How Africa Can Achieve the Sustainable Development Goals. Kigali: Sustainable Development Goal Centre for Africa (SDGC/A). 

\bibitem{sm} Kyakulumbye S, Pather S, Bagula A. (2020). Smart City Citizens’ Service Provision using Participatory Design and Participatory Sensing: Lessons for Developing Cities. In  IST-Africa Conference (IST-Africa) 2020 (pp. 1-12). IEEE.

\bibitem{who14} WHO (2014). The health of the people: what works. The African Regional Health Report 2014. Retrieved from: \url{apps.who.int/iris/bitstream/10665/137377/4/9789290232612.pdf}. Accessed 05/10/20
\end{thebibliography}
\end{document}